\documentclass[preprint]{aastex61}
\received{November 15, 2017}
\submitjournal{ApJS}

\renewcommand{\arraystretch}{1.5}

\begin{document}

\title{\textit{JINAbase} -- a database for chemical abundances of metal-poor stars}

\correspondingauthor{Anna Frebel}
\email{afrebel@mit.edu}

\author{Abdu Abohalima}
\author{Anna Frebel}
\affiliation{Department of Physics and Kavli Institute for
	Astrophysics and Space Research, Massachusetts Institute of Technology, Cambridge, MA, USA},
\affiliation{Joint Institute for Nuclear Astrophysics (JINA) -- Center for the Evolution of the Elements, East Lansing, MI, USA}

\begin{abstract}
Reconstructing the chemical evolution of the Milky Way is crucial for understanding the formation of stars, planets, and galaxies throughout cosmic time. Different studies associated with element production in the early universe and how elements are incorporated into gas and stars are necessary to piece together how the elements evolved. These include establishing chemical abundance trends, as set by metal-poor stars, comparing nucleosynthesis yield predictions with stellar abundance data, and theoretical modeling of chemical evolution. To aid these studies, we have collected chemical abundance measurements and other information such as stellar parameters, coordinates, magnitudes, and radial velocities, for extremely metal-poor stars from the literature. The database, \textit{JINAbase}, contains 1658 unique stars, 60\% of which have [Fe/H]$\leqslant-$2.5. This information is stored in an SQL database, together with a user-friendly queryable web application\footnote{http://jinabase.pythonanywhere.com}. Objects with unique chemical element signatures (e.g., r-process stars, s-process and CEMP stars) are labeled or can be classified as such. The web application enables fast selection of customized comparison samples from the literature for the aforementioned studies and many more. Using the multiple entries for three of the most well studied metal-poor stars, we evaluate systematic uncertainties of chemical abundances measurements. We provide a brief guide on the selection of chemical elements for model comparisons for non-spectroscopists who wish to learn about metal-poor stars and the details of chemical abundances measurements.

\end{abstract}


\section{Introduction} \label{sec:intro}

	Metal-poor stars, broadly defined, are stars with a heavy element content less than 1/10th that of the Sun \citep{ARAA}. They are generally classified as old Population\,II (Pop\,II) stars and predominantly reside in the halo of the Milky Way (e.g., \citealt{beers2000, carollo}), although a few stars have by now also been discovered in the bulge (e.g., \citealt{db_HOW15}). Satellite dwarf spheroidal galaxies also contain metal-poor stars (e.g., \citealt{cohen08}), and in particular, the ultra-faint dwarfs contain exclusively metal-poor stars (e.g., \citealt{frebel14}) given their short star formation and enrichment history. The most metal-poor stars are thought to have formed from gas only enriched by the very first stars, also called Population\,III stars (Pop\,III). These metal-free, massive short-lived stars are believed to have formed a few hundred million years after the Big Bang (e.g., \citealt{BROMM04,OSHEA07}) and were responsible for the first metal enrichments of the interstellar medium (ISM). This gave rise to the formation of the first Pop\,II stars (e.g., \citealt{GREIF10,WISE12,CROSBY13}) which contain the chemical signatures of these first supernovae. 

Thus, extremely metal-poor stars, defined here as stars with [Fe/H] $\leq -3.0$, have been the focus of many studies as they are an important tool for reconstructing the early chemical history of the universe. 
Particularly in the last decade, several large scale Galactic surveys, such as SDSS, SkyMapper, LAMOST, have been carried out leading to a rapidly increasing number of known extremely metal-poor  stars.
	
	Given that stars preserve the chemical composition of their birth gas cloud in their atmosphere, extremely metal-poor stars provide a unique record of the physical and chemical state of the early universe. In addition, kinematic analysis of their Galactic orbits yields clues to their origin, e.g., from accreted dwarf galaxies. Thus, by studying the chemical abundances of stars with different metallicities, Galactic chemical evolution can be mapped and understood \citep{MATTEUCCI12,NOMOTO13,MINCHEV14, cote17}. 
    
    Characteristic element signatures found in extremely metal-poor  stars can also enable the reconstruction of individual nucleosynthesis processes and events that occurred in the early universe \citep{HE1327_Nature,LUGARO08,HEGER10,ISHIGAKI14,ROEDERER_K16}. Knowledge of those provides critical aid for constraining chemical evolution models which is a highly complex subject. This also presents the intersection of observations and theory: observers provide metal-poor stars and their abundances and abundance patterns, and theorists use the nucleosynthesis predictions and chemical evolution models to reproduce the data to gain insight into how the Milky Way evolved. It is thus critical that observational results are made readily available for interpretation. In addition, small but important details pertaining to sample selection that are very familiar to observers, such as about data taking, analysis procedures, which lines, species and elements to use, relative vs. absolute abundances, etc., are often not known to theorists. This can lead to difficulties in using data as intended, and in the worst case, to misinterpretation of the data. 

While these challenges need to be individually addressed for each study at hand, we aim in this paper to describe some common ``pitfalls'' that theorists might want to consider when using abundance data of metal-poor stars. 
This stems an ongoing discussion in the Joint Institute for Nuclear Astrophysics (JINA) community and a desire for closer communication between observers and theorists. To aid in this process, we here present a web-based queryable database called \textit{JINAbase} that enables interested users to readily use abundance data from metal-poor stars in the literature. Alongside, we provide commentary on the many different element abundances and specific abundance signatures found in metal-poor stars to guide usage of the output of \textit{JINAbase}. 

As of July 2016, \textit{JINAbase} contains entries of 981 unique metal-poor stars with $\mbox{[Fe/H]} \le -2.5$ from the bulge, the halo, classical spheroidal dwarf and ultra-faint dwarf galaxies. The abundance and stellar parameter data collected is based on high-resolution (resolving power of $R = \lambda / \Delta \lambda \gtrsim 15,000$, with the majority having $R = 30,000 \,-\,40,000$) spectroscopic studies found in the literature. New results from forthcoming papers can be added by the authors after registering with the web app. Unfortunately, \mbox{\textit{JINAbase}} cannot offer a homogenized data set, but is simply a collection of literature results. 
Nevertheless, repeat analyses of many stars will allow, to some extent, a quantification of differences between various studies, and the field as a whole. Overall, it follows in spirit what was presented in \citet{FRE10LIST}, and is similar to the SAGA database \citep{SUD08}. However, \textit{JINAbase}, offers somewhat different functionality that includes labels, classifications and selection options to put together custom samples as described below.

The paper is organized as follows. In Section~\ref{sec:content}, we describe the literature data collected, what can be downloaded from \textit{JINAbase} for individual stars and samples. We also describe the web application's interface and how to query it. In Section~\ref{comparison}, we present a brief analysis on the multiple abundance sets for three well-studied metal-poor stars. We provide comments on various elemental abundances in order to assist choosing customized stellar abundance samples for comparison with model results in Section~\ref{samples}. We summarize in Section~\ref{summary}.

%========================================================
\section{Retrieving chemical abundances from \textit{JINAbase}}\label{sec:content}

	\textit{JINAbase} was constructed to provide easy and (near) complete access to metal-poor star abundances for the JINA and larger astronomy communities for a variety of projects that require stellar abundances of metal-poor stars for comparison with model predictions or other observational data. We have designed a queryable web application\footnote{The web application can be accessed at \href{http://jinabase.pythonanywhere.com}{jinabase.pythonanywhere.com}} using Python. The web application includes a user friendly interface enabling easy plotting of various abundances for a sample of stars, with many useful selection criteria,  which types of stars to be in- or excluded in the sample, and other features. 
     
    In this section, we describe which data and from where it was collected. We also describe the  interface of the web application and the general functionality of \textit{JINAbase}, how to use it to plot and extract stellar abundances, and how to query it.

 %===================================================   
	\subsection{Literature samples included in \textit{JINAbase}}
    
   We collected chemical abundances and stellar parameters for metal-poor stars (primarily with $\mbox{[Fe/H]} \leqslant -2.5$) from the period between 1991 to 2016. All references with links to the original papers and their bibtex entries are provided in the web application for easy access. These stars are located in the bulge, the halo, classical spheroidal dwarf and ultra-faint dwarf galaxies, and are classified as such (although some higher metallicity thick disk stars are included in the halo sample and not separately labeled). As of July 2016, \textit{JINAbase} contains data of 1658 unique stars with $\mbox{[Fe/H]}<0$ (with the lowest value being $\mbox{[Fe/H]}<-7.3$). Given multiple analyses of many stars, a total of 2734 entries are available. HD122563, with 28 entries, is the most well studied metal-poor star, see Section~\ref{comparison} for more information. \textit{JINAbase} contains 856 unique red giants ($\sim 52\%$), 469 subgiants ($\sim 28 \%$), 54 horizontal branch stars ($\sim 3 \%$) and 279 near main-sequence turnoff (TO) stars ($\sim 17 \%$). Selection criteria are further described in Section~\ref{sec:description}. These percentages show that observations of metal-poor stars are biased towards red giants, mainly due to the fact that they are observable out to larger distances than near main-sequence TO stars.  
       
\begin{figure}[t!]
    \centering
    \plotone{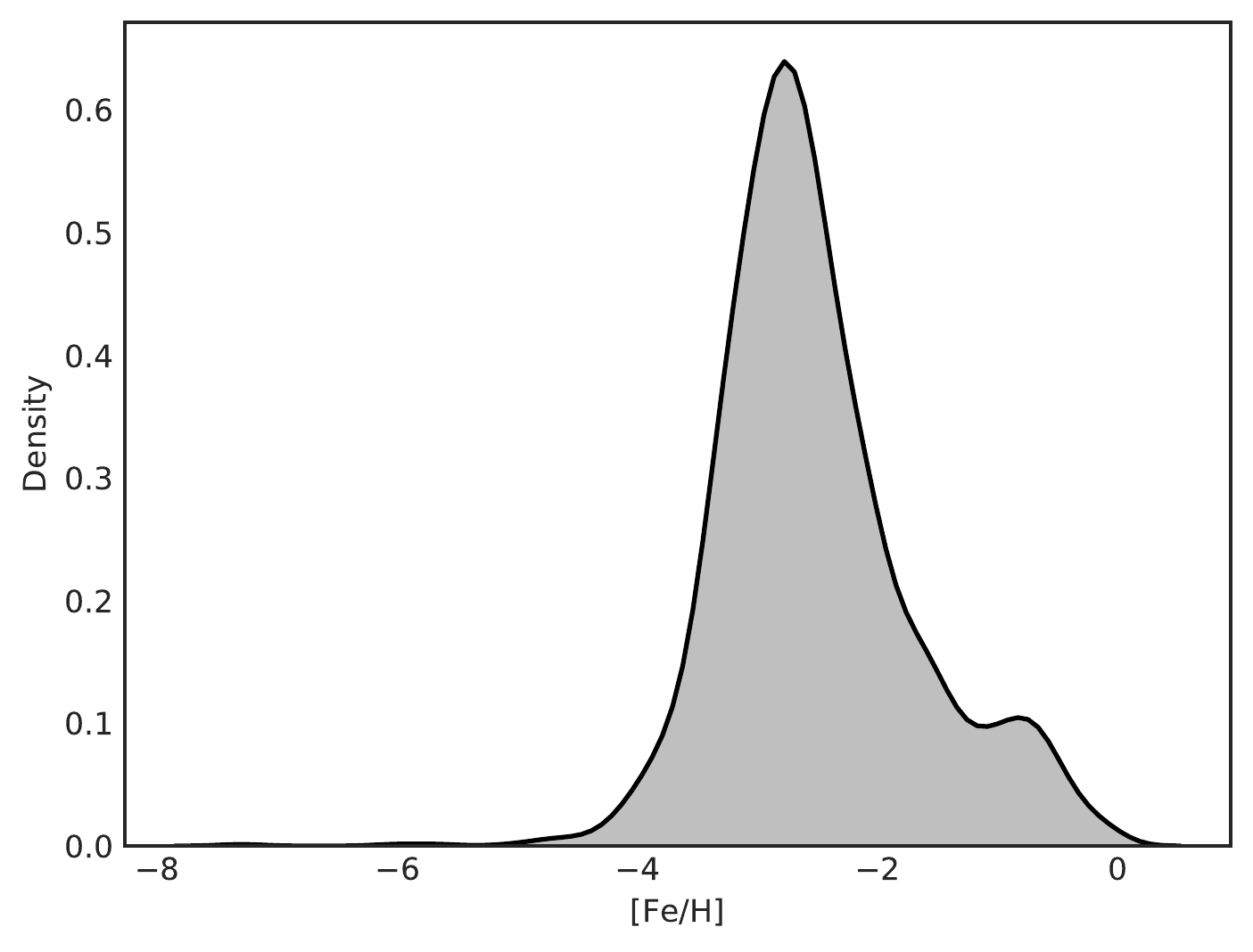}
    \caption{Metallicity distribution of the stars in \textit{JINAbase}, in the form of a kernel density estimation to visualize the metallicity range covered by the database content. When multiple entries for a given star are available, we chose the one with the largest number of measured abundances; see text for discussion. The main peak in the distribution is at $\mbox{[Fe/H]}=-2.75$. Any shape for $\mbox{[Fe/H]}>-2.5$ is likely not physically representative, as the database is increasingly incomplete there. There are two minuscule bumps in the distribution at $\mbox{[Fe/H]}\sim-5.5$ and $\mbox{[Fe/H]}=-7.3$, which the latter actually being nominally included to reflect a star with $\mbox{[Fe/H]}<-7.3$. \label{hist}}
    \end{figure}

       		Figure~\ref{hist} shows the density kernel smoothed metallicity distribution of the entire sample.
		For guidance at higher metallicity, i.e., for $\mbox{[Fe/H]} >-2$, we also added several samples of more metal-rich stars from the thick disk, such as the sample of \citet{db_FUL00}. In \textit{JINAbase}, they are part of the group halo stars and currently not separately labeled. However, their higher [Fe/H] abundances makes them easily identifiable.

       The steep decline towards lower metallicities naturally results from these stars being rare. The almost complete cutoff below $\mbox{[Fe/H]} = -4$ illustrates the rarity of the most metal-poor stars. Only 22 unique stars have $\mbox{[Fe/H]} \leqslant -4.0$, 7 of which have $\mbox{[Fe/H]} \leqslant -4.5$, compared to a total of 981 unique stars with $\mbox{[Fe/H]} \leqslant -2.5$. The sharp drop off towards higher metallicities is due to selection effects in this literature compilation, as we focused on collecting data for stars with $\mbox{[Fe/H]} \leqslant -2$. Consequently, above $\mbox{[Fe/H]}=-2$, studies are lacking, and in addition, not all stars discovered may have (published) high-resolution abundances. The \textit{JINAbase} upload feature enables incorporating additional stellar abundances and future results, e.g., from higher-metallicity stars or globular clusters.	
        
We also added information on each star taken from the SIMBAD database\footnote{http://simbad.u-strasbg.fr/simbad/} and Vizier catalogs\footnote{http://vizier.u-strasbg.fr/viz-bin/VizieR}. This way, coordinates, magnitudes and radial velocities can be readily extracted from \textit{JINAbase} together with the stellar parameter and element abundance data. While these are interesting and important quantities that round out the information content of \textit{JINAbase}, we note that no quality checks or source inspection was done. This should be kept in mind when using the auxiliary information. For example, the radial velocity values listed in SIMBAD may not be the most accurate or latest ones available. 
		
	Usually many different designations exist for a given object, in \textit{JINAbase} we try to give each star a unique identifier to keep track of all entries per star. We identify each star in the SIMBAD database, and then use the main SIMBAD designation as the distinct SIMBAD identifier in \textit{JINAbase}. This step is of great importance as 437 stars have multiple entries (see Section \ref{comparison}). The name used in the reference for the star is saved in \mbox{\textit{JINAbase}} as well.

       %=================================================================
\subsection{What is \textit{not} included in \textit{JINAbase} and cautionary notes}        

The primary goal of \textit{JINAbase} is to provide chemical abundance data as obtained from the various studies in the literature. Accordingly, there naturally are limitations as to what \textit{JINAbase} can provide, as there are multiple intermediate measurements and steps individual to every study that lead to these abundances, such as equivalent width measurements and line lists used. None of these are included in the database (with the exception of the stellar parameters, see below) but can be obtained from the original papers. Neither included here are measurement uncertainties. Reasons are numerous and include that there is currently no homogeneous prescription for determining uncertainties; some studies adopt standard deviations as final uncertainties, others standard errors. Others include and/or assess systematic uncertainties. 

For studies wishing to employ abundance uncertainties on the data provided in \textit{JINAbase}, we recommend consulting with the original papers or to adopt typical uncertainties for high-resolution abundance analysis, as derived in Section~\ref{comparison} from repeat measurements of three metal-poor stars. Alternatively, for a proper and homogeneous assessment of abundance uncertainties, a homogeneous re-analysis of all or at least a large number of stars would need to be undertaken, along the lines of what was done in \citet{db_YON13}. Of course this also pertains to the actual chemical abundances that have  been determined in somewhat individual ways by each study. Original papers should always be consulted for future details. 

Furthermore, when collecting abundances we obtain the final or best set of abundances presented. We thus do not distinguish between LTE and NLTE abundances. But the vast majority reports LTE abundances and only some of the more recent studies have begun to provide NLTE stellar parameters and/or abundances. All these inhomogeneities also affect stellar parameters on which the abundance measurements are based. They have been determined by a variety of methods (e.g., purely spectroscopic, partially based on photometry; different line lists; different analysis codes; etc.) that have not been recorded here. But it is safe to assume that a range of systematic differences arise from using different stellar parameter methods.

Finally, the completeness of this literature compilation is nearly impossible to assess, given all the different samples. Overall, the best guess would be that the completeness level increases with decreasing [Fe/H], given the importance of these stars, and regardless of their location (i.e., halo, dwarf galaxy). Assuming that all stars known below $\mbox{[Fe/H]} \leqslant-3.5$ have published high-resolution abundances available (which, again, may or may not be true), we speculate that \textit{JINAbase} is complete for these stars. For halo stars with higher [Fe/H], given all the recent survey results, it is clear that this sample is (increasingly) incomplete. This must be taken into account when attempting to reproduce the metallicity distribution function. For dwarf spheroidal galaxies, the situation is similar to that of the halo, with the sample being likely fairly complete up to the $\mbox{[Fe/H]} \leqslant -2$ level and down to a certain magnitude. For ultra-faint dwarf galaxy stars, the situation is likely better. Samples are naturally magnitude limited given the constraints of current telescopes, so that these samples should be fairly complete.  In any case, original studies should be consulted to learn more about completeness levels of individual samples.

All in all, the truth is that the entire body of chemical abundances of metal-poor stars is not as homogeneous as one would desire for comparison with e.g. chemical evolution models. Nevertheless, selecting specific samples, working with abundance ratios, understanding selection effects and awareness of the associated systematic uncertainties can somewhat alleviate this issue. The goal of this paper is to provide this sort of guidance.

 %=================================================================
 
    \bgroup
    \def\arraystretch{1.3}
      \begin{deluxetable*}{l l}
      \tablecaption{\textit{JINAbase} content description\label{tab:contents}}
      \tablehead{\colhead{Data category} & \colhead{Description}}
      \startdata
       {Labels assigned} & \textit{JINAbase} ID number\\
       {by \textit{JINAbase}} & Priority label: to choose from stars with multiple entries\\
			& Evolutionary status of star: RG, HB, SG, or MS\\
			& Label for carbon enhancement: CEMP, or CEMP-no\\
        \colrule
		{Literature} & Reference code that associates original paper \\
        \colrule
		{Identifiers} & Star name from the respective paper \\
			& SIMBAD identifier \\
		\colrule
		Labels assigned & Label for neutron-capture element enhancement: r-I, r-II, s-rich, i-rich, or r+s stars\\
		by authors& Location: Halo\,(HA), bulge\,(BU), ultra-faint dwarf galaxy\,(UF), or classical dwarf galaxy\,(DW)\\
		\colrule
		{Position} &Right ascension (J2000) (from SIMBAD)\\
			& Declination (J2000) (from SIMBAD)\\
     	\colrule
		{Radial Velocity} & Radial velocity (from SIMBAD)\\
			& Reference for radial velocity (from SIMBAD)\\
		\colrule
		{Magnitudes} & Values for U, B, V, R, I, J, H, K (from SIMBAD)\\
        \colrule
        Stellar parameters &  Effective temperature, surface gravity, metallicity, microturbulence (from respective paper)\\
		\colrule
        {Chemical} & $\log\epsilon(\rm X)$, abundances for elements from Li to U, when available\\
        abundances& $\log\epsilon (\rm X)$ values for elements with two ionization states (Ca, Ti, Cr, Mn, Fe), when available\\
        \enddata
%      \tablecomments{}
%

      \end{deluxetable*}
    \egroup

	\subsection{\textit{JINAbase} content}\label{sec:description}
    
        An overview of the complete content of \textit{JINAbase} can be found in Table~\ref{tab:contents}. All or parts of the table can be downloaded as an ascii file, according to chosen selection criteria. In the following, we describe all available information in more detail. 
        
        \smallskip
        \textbf{Labels assigned by \textit{JINAbase}:} \textit{JINAbase} assigns an internal ID number for tracking stars. This ID is needed, for example, when plotting abundances versus atomic number. 
                
          Following \citet{FRE10LIST}, \textit{JINAbase} assigns a priority label to offer the option of plotting a sample in which each point represents one unique star. This is relevant for stars studied by multiple authors, e.g., HD122563. The priority label ``1'' is given to the study with the most measured abundances. Stars with just one entry are also assigned priority ``1''. A rough assumption is that a study with more elements reported likely had good or better data than other studies. In the individual case this is of course not true, but here, it assists making a simple choice for users to avoid duplicate plotting. Nevertheless, the user has full control over switching this feature on or off. 

		Based on stellar parameters collected from the literature and a 12 Gyr isochrone \citep{Y2_iso}, \textit{JINAbase} assigns a label for the  evolutionary status of the star. The following groups are distinguished:\\
\textbf{Red giants:} \\
		For $\mbox{[Fe/H]} \geqslant -2$: $T_{\rm eff} \leqslant 5400$\,K and $\log g \leqslant 3.5$\\
		For $\mbox{[Fe/H]} \leqslant -2$: $T_{\rm eff} \leqslant 5600$\,K and $\log g \leqslant 3.5$\\
		\textbf{Subgiants:}\\ 
		For $\mbox{[Fe/H]}\geqslant-2$: $6700\ge T_{\rm eff}\geqslant 5400$\,K, $4\ge\log g\geqslant3.5$\\
		For $\mbox{[Fe/H]} \leqslant -2$: $6700 \ge T_{\rm eff} \geqslant 5600$\,K, $4 \ge \log g \geqslant 3.5$
		\textbf{Horizontal branch stars:}  $T_{\rm eff} \geqslant 5400$\,K, $\log g \leqslant 3.5$\\
		\textbf{Main-sequence TO stars:} $\log g > 4$, for all [Fe/H] and $T_{\rm eff}$
          
        \smallskip
          In addition, \textit{JINAbase} assigns labels to carbon-enhanced metal-poor (CEMP) stars. CEMP stars (with enhancement in neutron-capture element abundances) and CEMP-no stars (with \textit{no} enhancement in neutron-capture element abundances) are distinguished, when carbon and barium abundances are available. The physical meaning of these labels is yet to be clarified. For the time being they serve to classify metal-poor stars. We adopted the definition from \citet{db_AOK07a} for carbon-enhanced stars: \\ 
		\textbf{CEMP:} $\mbox{[C/Fe]} > 0.7$\\
		\textbf{CEMP-no:} $\mbox{[C/Fe]} > 0.7$ and $\mbox{[Ba/Fe]} < 0$ 
    
%        Other paths to assigning priorities are currently being explored.

\smallskip
        		\textbf{Literature:} \textit{JINAbase} assigns a reference code that consists of the first three letters of the name of the first author together with the year of publication, e.g. FRE10.
                 
        \smallskip
               	\textbf{Identifiers:} There are two designations for a given star. The one provided by authors of the abundance study, and the one that is the primary SIMBAD identifier (although in some cases, they will be the same). \textit{JINAbase} records both.

        \smallskip
\textbf{Labels to be assigned by authors:} When uploading new abundance results, authors are asked to assign certain science labels to their stars if they show characteristic abundance signatures. For a classification as r-I, r-II, s-rich, i-rich, or r+s rich star, the abundance pattern of neutron-capture elements should follow the respective pattern, for unambiguous identification. We request this in addition to the abundance ratio criteria given below, where we adopted the definition from \citet{ARAA}, and Frebel 2018 (in prep.) 
for neutron-capture element-enhanced stars:

\smallskip
\noindent \textbf{r\,-I:} $0.3\leqslant\mbox{[Eu/Fe]}\leqslant1$ and $\mbox{[Ba/Eu]}<0$\\
		\textbf{r\,-II:} $\mbox{[Eu/Fe]} > 1$ and $\mbox{[Ba/Eu]} < 0$\\
		\textbf{s:} $\mbox{[Ba/Fe]} > 1$ and $\mbox{[Ba/Eu]} >0.5$\\
		\textbf{i:} $0 < \mbox{[Ba/Eu]} < 0.5$ (formally ``r/s'' stars; see \citealt{ARAA})\\
        \textbf{r+s:} $\mbox{[Ba/Fe]} \approx 1.0 $ and $\mbox{[Ba/Eu]}\approx0.5$. 
        \smallskip
        
        The last classification is new following the discovery of the first star that unambiguously shows a combined chemical pattern from the r-process and the s-process \citep{gull17}. Hence, the \textquotedblleft r+s" label. As more r+s stars are being discovered, this criterion may need to be adjusted; it currently assumes a minimum r-process contribution of $\mbox{[r/Fe]}\sim0.3$\,dex (Frebel 2018, in prep.). It is slightly different from the  \textquotedblleft r/s" classification of \citet{ARAA} -- the abundance patterns of these stars are likely having a different origin, namely in the i-process \citep{hampel16}. More details on these groups is given below.

        \smallskip
        Additional science labels may be introduced in the future.

\smallskip
Authors are also asked to assign a location label to their stars, such as halo, bulge, ultra-faint dwarf galaxy, or classical dwarf galaxy. This assists in characterizing all Galactic stellar populations as well as the classical and ultra-faint dwarf galaxies that host e.g., extremely metal-poor  stars and when comparing stellar abundances of different populations.

		We distinguish four main locations for metal-poor stars:\\
		\textbf{Halo:} The halo of the Galaxy hosts about 90\% of all the unique stars included in \textit{JINAbase} since we deliberately included a few thick disk metal-rich stars for comparison. Among known metal-poor stars with $\mbox{[Fe/H]} \leqslant -3$, 394 out of 416 stars are members of the field halo population of the Galaxy, the remaining 22 stars are located in dwarf galaxies and the bulge. The high halo fraction is mainly due to the observational advantages with regards to stars in the halo. The halo is on average metal-poor and it is sparsely populated allowing for clear lines of sight towards the stars, compared to e.g. the dense bulge region.\\
		\textbf{Bulge:} Recent efforts to find metal-poor stars in the bulge have added 45 stars to the general population of metal-poor stars.\\
		\textbf{Classical dwarf galaxies:} 53
        %% [xx we are including more atm, WORK IN PROGRESS]
        %% not in jinabase: Leo\,T \& II, Canes Venatici\,I,
        stars in \textit{JINAbase} are located in the classical dwarf galaxies such as Draco, Ursa Minor, Carina, Sextans, Leo\,I, Sculptor, and Fornax.  \\
		\textbf{Ultra-faint dwarf galaxies:}
        %% not in jinabse: Tucana\,III, Ursa Major\,I
        There are 43 stars identified as members of ultra-faint dwarf galaxies. This includes the following galaxies, Segue\,1 \& 2, Reticulum\,II,  Coma Berenices,  Canes Venatici\,II, Leo\,IV, Ursa Major\,II, Bootes\,I \& II, and Hercules. \\
        
       \smallskip
  	\textbf{Position:} Coordinates right ascension and declination (J2000) are collected from SIMBAD, following the identification of the star in the SIMBAD database.

	\textbf{Velocity:} Available radial velocity measurements and references collected from SIMBAD. No vetting has been applied; they might not be the latest values or could be based on medium-resolution spectroscopy rather than high-resolution spectroscopy. The corresponding references should be consulted for more information. 
                        
	\textbf{Magnitudes:} Available magnitudes (U, B, V, R, I, J, H, K) are collected from SIMBAD
                       	
	\textbf{Stellar parameters:} Effective temperature, surface gravity, metallicity, microturbulence are collected from the same references as the abundances.

	\textbf{Chemical abundances:} $\log \epsilon(\rm{X})$ abundances, for elements from Li to U, when available. Abundances for some elements with two ionization states are also included.\\

%===========================================

\subsection{Comments on the element signature labels}\label{comm}

As described above, labels are assigned to stars with characteristic abundance signatures for an easy selection of these stars to test nucleosynthesis and chemical evolution models. Here we provide additional comments on these groups of stars. However, we note that all of the labels are completely optional for users to make use of. Also, these labels cannot replace a thorough understanding of the abundance data and their pedigree.

%\textbf{$\alpha$-enhancement:} Most halo stars show enhanced abundances in Mg, Ti, Ca, Si ($\alpha$-elements). The ``standard'' enhancement of [$\alpha$/Fe]\,$=$\,0.4 has been interpreted as a signature of core-collapse supernovae. Some EMP stars, however, have very low levels of $\alpha$-elements, and even subsolar abundances. This could be a signature of the first supernova type Ia (providing iron and no $\alpha$-elements) exploding already in the early universe. This abundance enhancement is not assigned a signature label in the database.
	
	\textbf{Carbon-Enhanced Metal-Poor stars}: 
    
    When constructing a sample of carbon-rich stars for any model comparisons, the scientific question drives sample selection. For studies of the carbon enrichment of early gas, s-process stars, i-process stars and r+s stars have to be strictly excluded (by deselecting the relevant boxes) as the carbon abundances of these stars are not representative of the stars birth environment and instead in all likelihood from mass transfer events from binary companion stars. On the contrary, for studies, e.g. related to the dredge up of nucleosynthesis products, or mass transfer processes, or early asymptotic giant branch (AGB) carbon production, then a sample of s-process or i-process stars needs to be chosen. 
    
    Overall, $25\%$ of observed extremely metal-poor stars with $\mbox{[Fe/H]} \leqslant -3.0$ are carbon-rich with $\mbox{[C/Fe]}>0.7$. However, as stars are ascending the giant branch, carbon gets converted to nitrogen. Observed carbon abundances thus do not necessarily reflect the composition of the birth gas cloud, i.e., they show signs of self-enrichment in C and N. \citep{PLACCO14} calculated such corrections using stellar evolution models. This enables to study the birth composition of these stars. This then also allows to determine the true frequency of CEMP stars upon exclusion of s-process, i-process and r+s stars whose large carbon abundances are not arising from the birth cloud but from mass transfer events which masks any contribution from self-enrichment along the giant branch. Knowing the true frequency can assist in better understanding the many origins of carbon, which remain to be understood in detail as sources seem ubiquitous. Options include different types of early supernovae that pre-enrich the gas from which the extremely metal-poor stars form. This is of particular importance, as the fraction of CEMP stars steadily increases with decreasing metallicity, from $25\%$ at $\mbox{[Fe/H]} \leqslant -3.0$ up to 100\% at the lowest [Fe/H]. 
    
	\textbf{CEMP-no stars:} These are CEMP stars that show no overabundance (i.e., ``normal'', subsolar levels) in neutron-capture elements, as measured by barium. Most CEMP stars with $\mbox{[Fe/H]}<-3.0$ are CEMP-no stars, as s-process stars begin to rise only at $\mbox{[Fe/H]}>-2.6$ \citep{simmerer04}. The lack of neutron-capture element enhancements at the lowest metallicities remains to be understood but observationally establishing their frequency will help address any underlying enrichment processes.
	
	\textbf{r-process:} About 5\% of metal-poor stars with $\mbox{[Fe/H]}<-2.5$ show a strong enhancement (r-II stars) in r-process elements \citep{db_BAR05}. Another 15-20\% show a mild enhancement (r-I stars). Their chemical signature of heavy elements above barium follows the scaled solar r-process pattern\footnote{The scaled solar r-process pattern is obtained from subtracting the theoretically well understood s-process component from the total solar abundances. It is thus a derived product, not a measurement.}. Among light neutron-capture elements, there are variations with respect to the scaled solar pattern, though. Reasons for this are largely unknown but include possibilities for multiple r-processes production sites \citep{Hilletal:2002}. Presumably, these stars formed from gas that was enriched in these elements by a neutron star merger \citep{db_JI16a} or an unusual jet-drive supernova \citep{Winteler12} in the early universe. Supernovae also contribute light neutron-capture elements (from strontium to barium) in a limited r-process \citep{Wanajo13}

	\textbf{s-process stars:} Some metal-poor stars display large amounts of neutron-capture elements associated with the s-process in their spectra. These stars are thought to be in binary systems. A former primary AGB companion transferred mass to the (now observed) lower mass companion, enriching it with s-process elements as well as large amounts of carbon that were initially produced in the AGB star \citep{karakas10,BISTERZO12}. This is why large amounts of carbon (e.g., $\mbox{[C/Fe]} >1.0$) are always present in these stars. Today only the lower mass secondary star is observed. It should be noted here that AGB stars are numerous throughout the universe, and thus significantly enrich their surrounding ISM with carbon and s-process elements through their stellar winds in their own right. This leads to significant contributions to the global chemical evolution, and the carbon and s-process element enrichment of the gas from which all subsequent (metal-poor) stars formed. Such global enrichment appears to occur only from $\mbox{[Fe/H]}>-2.6$ \citep{simmerer04}, due to the delay time needed for the very first lower mass stars to become the first AGB stars in the universe. This enrichment channel has to be included in chemical evolution models, whereas s-process metal-poor stars that receives their chemical signature from a mass transfer event, provide tests for AGB nucleosynthesis, carbon production, and binary systems. 

	\textbf{i-process stars:} Some metal-poor stars show strong enhancements in neutron-capture elements that do not match either the scaled solar s- or r-process. These stars were termed ``r+s'' stars because it initially looked like a combination of these two processes would explain the observations. But many studies showed that another explanation is needed for these stars \citep{db_JON06}. Recent studies have then invoked the ``intermediate'' (i-)process that may also occur in AGB stars \citep{hampel16,denissenkov17}. While more studies are needed to fully explain these observed patterns, it seems that the i-process is likely responsible for these neutron-capture signatures. All i-process metal-poor stars would then be in binary systems which received enriched material from their AGB companion. 
    
    \textbf{r+s stars:} So far, only one star has been found with a neutron-capture element abundance pattern that can plausibly be explained with a combination of the r- and s-process patterns \citep{gull17}. The star likely formed from gas enriched by an r-process and then later received s-process material from a companion in a binary system. Future searches will hopefully uncover more of these stars. 

\smallskip	
	In closing, we note that the carbon enhancement naturally goes along with most of these signatures, except for r-process stars due to the  (extrinsic) enrichment of the gas from which they formed. Nevertheless, a few carbon-enhanced r-process stars are known, e.g., CS22892-052 \citep{db_SNE03}. Looking forward, observationally establishing the frequencies of each group of stars will help to gain a deeper understanding of their astrophysical site of production. In turn, theoretical modeling will be needed to learn more about the associated nucleosynthesis processes.

	\subsection{The \textit{JINAbase} web application} 
    
		The web application is divided into tabs. A navigation bar includes the four main tabs; Home, Query/Plot, Search and References. Additionally, the navigation bar includes the option for users to log in or register. Users can register to gain access to the web application's advanced functionalities. The registered user can upload data to the database and edit pre-existing data (see also Section~\ref{usage}). We added this feature to facilitate maintaining the database, this way it is a community effort. The \textbf{Home} tab includes a brief general description of the tabs and lists recent updates from the developer. It also includes information for how to cite the database, as well as contact information to report bugs, and suggestions for further improvement.
        
\begin{figure*}
        	\centering
            \plotone{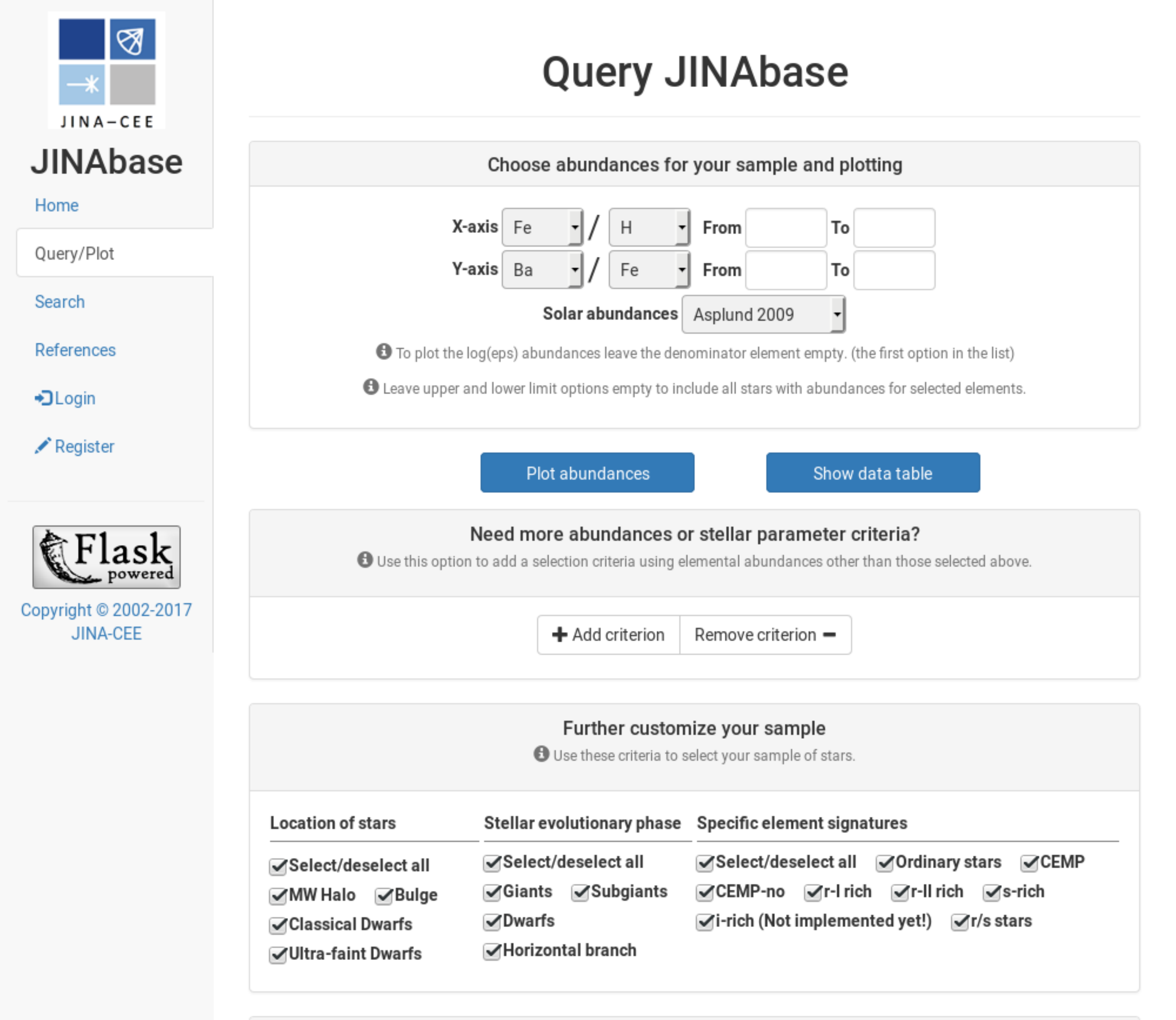}
            \caption{A screen shot of the Query/Plot tab from the web application. This figure shows the panels displayed in the top half of the tab. These panels guide the user through selecting their sample of metal-poor stars. The remaining panels are shown in Fig~\ref{fig:webapp-plot-2}. \label{fig:webapp-plot}}
		\end{figure*}

\begin{figure*}
        	\centering
            \plotone{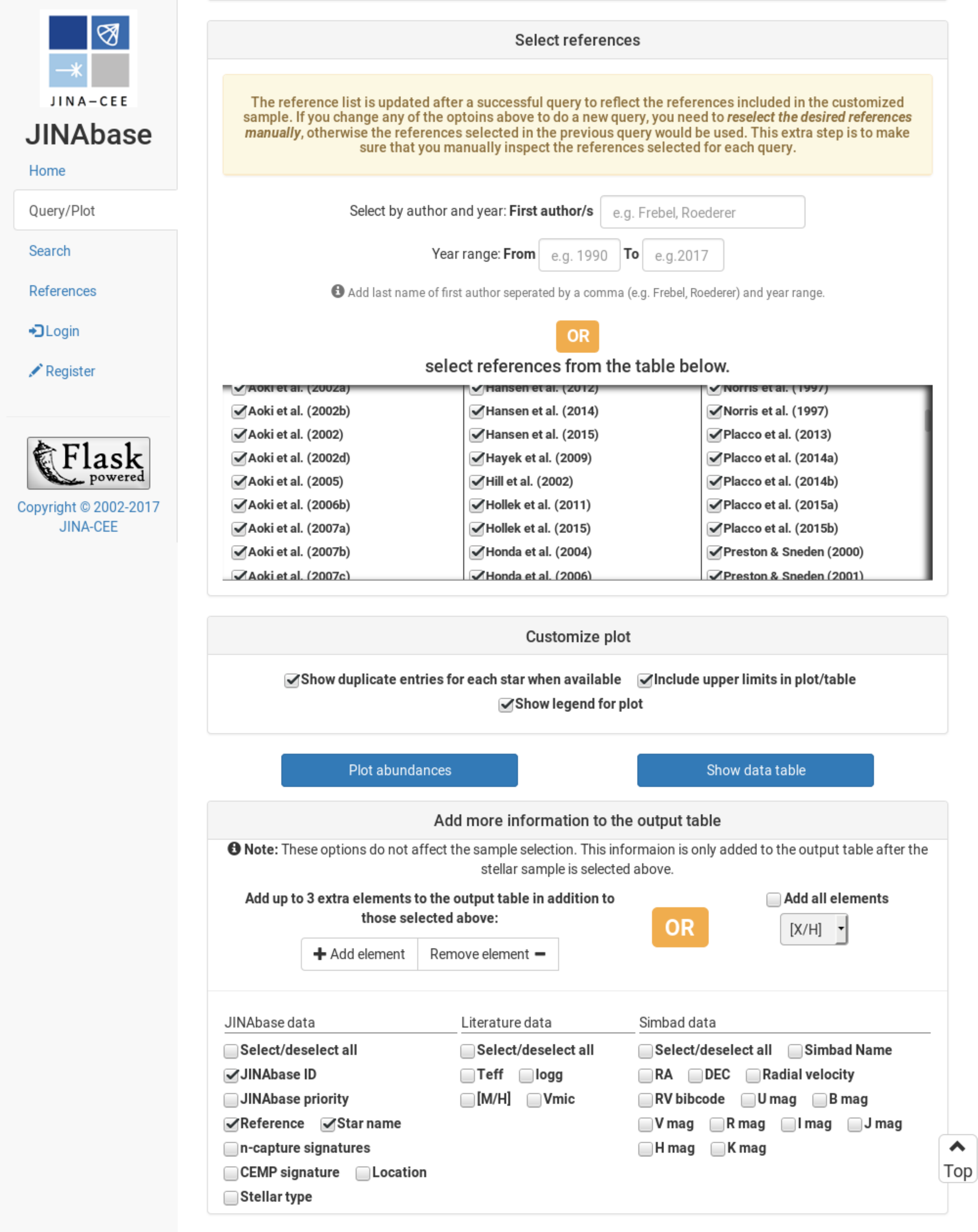}
            \caption{The second half of the \textbf{Query/Plot} tab from the web application. \label{fig:webapp-plot-2}}
		\end{figure*}

        The \textbf{Query/Plot} tab interface is divided into several panels guiding the user through the steps needed to choose a particular sample of stars. A screen shot of the web page is shown in Fig~\ref{fig:webapp-plot},\ref{fig:webapp-plot-2}. The first panel has the options to select the main abundances/stellar parameters to query or plot as well as which solar abundances to use if needed. The selection is divided into x and y-axes data, with a nominator and denominator selection for each. These options form the base for any selected sample. For ease of plotting, empty ``From'' and ``To'' boxes plot all available data. Another option is to define either the ``From'' or ``To'' box alone, leaving the other box empty to set no limit. The user can also add additional selection criteria to customize the sample using other chemical elements or stellar parameters (T$_{eff}$, [Fe/H], $\log g$, v$_{t}$), this feature is available in the second panel. Up to three extra criteria can be added. This way, the user can choose, e.g., user defined carbon-enhanced stars or select only stars in a specific effective temperature range.

      Next, there are a number of choices the user can make to further refine the sample selection. One can choose from locations of the stars (Milky Way halo, bulge, classical dwarfs and ultra-faint dwarfs), stellar evolutionary status (red giants, subgiants, main-sequence (near) turnoff, and horizontal branch stars) and pre-defined characteristic element signatures (ordinary stars with no special element signatures, r-I and r-II r-process, s-process, i-process, r+s-process stars, as well as CEMP and CEMP-no stars). The selection criteria for these groups of stars are described in Section~\ref{sec:description} and further comments are given in Section~\ref{comm}.
        
   At this point the sample of stars desired is defined. Then, all or any set of references can be selected to query. Selecting at least one references is required to query the database. All or individual references can be selected from the provided table or a first author's last name (separated by a comma if more than one) and/or a year range can be entered in the search boxes to query the list of references included in \textit{JINAbase}. After the references are selected, the user can plot the selected sample to explore it using an interactive plot. We recommend exploring the sample through the plot before generating the output table.
   
    Finally, there are options for adding supplementary information to the output table to download. First, when attempting to extract data for a specific sample, the user can add up to three extra elements to be included in the output table, otherwise just the two entries selected at the very top for the x and y-axes will be outputted. Alternatively, \textit{all} available abundances for the selected sample can be outputted when ticking the respective box.
    
    In addition, the user can specify additional information. This includes literature stellar parameters (which are used e.g., above to calculate the stellar evolutionary status), additional \textit{JINAbase} information (\textit{JINAbase} ID, \textit{JINAbase} priority key (based on the study with largest number of chemical abundances), literature star name) and SIMBAD information (SIMBAD name, coordinates RA DEC, different magnitudes and radial velocity information).
       
%====================================================       
        Users can search \textit{JINAbase} for a star (list of stars) using the \textbf{Search} tab. Similar to the \textbf{Query/Plot} tab, this tab is divided into panels. There are three main options for searching in \textit{JINAbase}; 1)using the literature or SIMBAD identifier or by RA \& DEC coordinates given a search radius, 2)search by reference(s) using the references table, 3)search using \textit{JINAbase's} ID, if known before hand (e.g. from the \textbf{Query} tab). These three options are independent, priority is given to each option by its order.
        
        There are two options to view the result of a search, either by plotting the chemical abundances (specifying the desired format) versus the atomic number or as a table. The format and atomic number range of the chemical abundances can be specified from the second panel, along with the solar abundances to use if needed. The result table includes, by default, the chemical abundances of the star(s). The user can add extra information to the table using the given options in the last panel, just like in the \textbf{Query/Plot} tab. The user can then save the table as a simple ascii file.
        
        \begin{figure*}
        	\centering
            \epsscale{0.8}
            \plotone{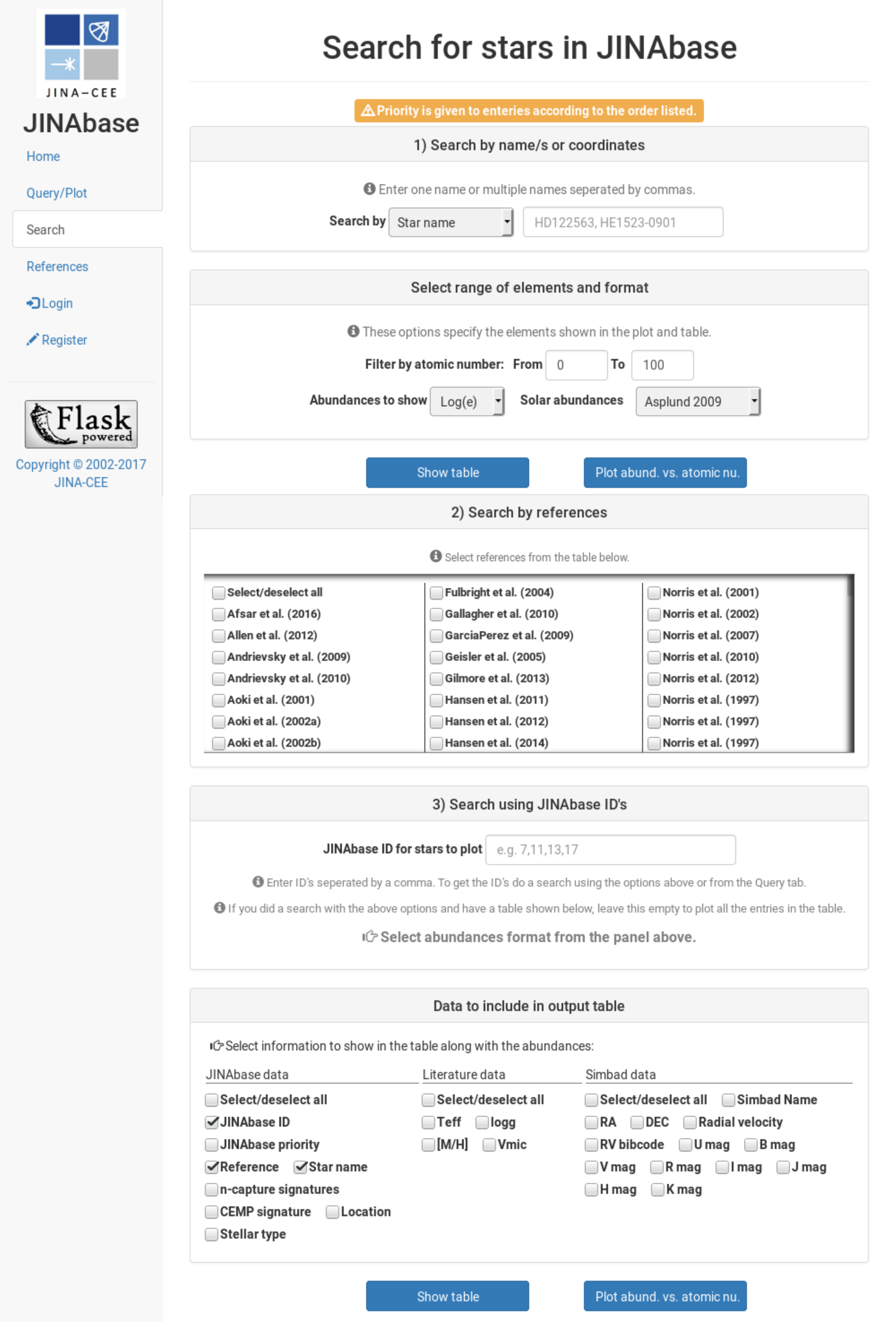}
            \caption{A screen shot of the \textbf{Search} tab from the web application. \label{fig:webapp-search}}
		\end{figure*}
        
        The \textbf{References} tab includes a table with all the literature included in the database. Links to the papers and the bibtex entries on ADS are included, as well as the number of stars per study. Each entry in the output tables is assigned the corresponding reference code on this page. This enables easy referencing of all original papers after constructing custom samples. Extra options are added to the navigation bar depending on the role assigned to the logged in user.

%======================================

 \subsection{How to extract \textit{JINAbase} content -- examples}

We now comment on and describe several examples for common questions and queries.

	\textbf{What additional data products are available for download?} Abundances [X/H] and [X/Fe] are calculated with user-chosen solar abundances. They are used for plotting and can also be downloaded.

\textbf{How to extract stars below a specific [Fe/H]:}
	Plotting (and downloading) stars below a certain [Fe/H] value is  one of the most common requests. In \textit{JINAbase}, there are two ways to achieve this. 
    
\smallskip
	1) Select stars based on the metallicity given as part of the stellar parameters. (In many cases, the overall metallicity (or model metallicity) employed to construct the model atmosphere differs from the iron abundance ultimately calculated with said model atmosphere). A sample of stars below a certain model metallicity value can be selected by choosing from the dropdown menu [Fe/H] for the x-axis ``nominator element'' and leaving the ``denominator element'' empty (the first option in the list). Alternatively, iron and hydrogen could of course also be specified to impose an [Fe/H]-based selection. This would select on the calculated [Fe/H] abundances. Then choose another stellar parameter from the dropdown menu for the y-axis ``nominator element'' and leaving the ``denominator element'' empty. This ensures that all stars with [Fe/H] have a matching y-axis value. (The user could choose any element there but then only stars which have this element abundance available will be selected. But stellar parameters are available for all stars). Then specify either the \textbf{From} or \textbf{To} values. Any blank box means that no limit of the data is placed; everything will be included. But both the \textbf{From} and \textbf{To} values can also be specified, depending on need. Make sure no additional abundance criteria are selected from the second panel. To include all abundances of all stars in the output table, tick the box at the bottom of the page to \textbf{Add all elements}, specifying the preferred format. Then \textbf{Plot abundances} or \textbf{Show data table}. The data table can then be downloaded (in ascii format). (Don't forget to also choose solar abundances, location of stars, evolutionary phases and specific element signatures). 

	\smallskip
	2) A sample of stars below a certain model metallicity value (see above) can be selected by adding a user defined criterion and choosing \textbf[Fe/H] from the dropdown menu of the ``nominator element'' and leaving the ``denominator element'' empty or ignoring its value). This selection is in addition to the elements selected above for the x and y-axes, i.e., only stars will be shown that have those two element abundances available \textit{and} adhere to the specified [Fe/H] cut. Alternatively, iron and hydrogen could of course also be specified to impose an [Fe/H] cut on the sample chosen above. Then, specify either \textbf{From} or \textbf{To}, or both (see above). Then \textbf{Plot abundances} or \textbf{Show data table}. The data table can then be downloaded. (Don't forget to also choose solar abundances, location of stars, evolutionary phases and specific element signatures).

\textbf{How to plot T$_{\rm eff}$ vs log\,\textit{g}:}
	In the \textit{Query/Plot} tab, select \textbf{T}$_{\rm eff}$ for the x-axis and \textbf{log\textit{g}} for the y-axis (as the ``nominator elements'' and leaving the ``denominator elements'' empty or ignoring them). Switch \textbf{From} and \textbf{To} values to force reverse plotting of the axes to obtain a Hertzsprung-Russell-Diagram plot. Add an [Fe/H] or other abundance constraint by choosing an \textbf{Add criterion}, if desired. 

\smallskip
\textbf{How to download the entire \textit{JINAbase} content:}
    The content of the database can be downloaded in its entirety using the following steps. However, we highly recommend only downloading subsamples from the database, as this option is currently not fully optimized and the displayed table is extremely laggy (due to the enormous size of the table). To retrieve all the information from the database, in the \textit{Query/Plot} tab, select any stellar parameter for the x-axis and any stellar parameter for the y-axis. Select all locations of stars, evolutionary phases, specific element signatures and references. Then, tick the box at the bottom of the page to \textbf{Add all elements}, selecting the preferred format. If needed select the solar abundances to use at the top of the page. Then \textbf{Show data table} to view the output table. The data table can then be downloaded. To include all or some of the supplementary information available, select those desired from the last panel at the bottom of the page.
    
\subsection{Using \textit{JINAbase} in your research now and in the future}\label{usage}
% zzz

The goal for constructing \textit{JINAbase} has been to provide easy access to the chemical abundance data of metal-poor stars. While this is largely achieved with the web-based queryable tool, we have also created a black box in the sense that analysis details are not stored but will in all likelihood be different for every study included. This means that there is no substitute for checking, and citing, the original papers. \textit{JINAbase} links to all papers for easy access, and also provides bibtex entries. If you used \textit{JINAbase} for your work, this paper (Abohalima \& Frebel 2018) should kindly be cited as well.
%% cite using \software command

Long-term, \textit{JINAbase} will be a \textquotedblleft living" database that is continuously updated as new results become available. Results from new (or currently missing) papers can be uploaded by authors themselves after signing up as contributors. This will assist in keeping the database content up to date. Registration and login links can be found in the navigation bar at the \textit{JINAbase} website. Uploading instructions are provided upon logging in. Alternatively, the authors of this paper can be notified about large data sets to include. 

There are also possibilities to later on include abundance data for e.g., stars in globular clusters, metal-rich stars from various populations and secondary information such as distances and kinematics from Gaia, when available. Large survey samples reporting chemical abundances from e.g. APOGEE\footnote{http://www.sdss.org/dr12/irspec/spectro\_data/} (\citealt{APOGEE16}) would also be helpful to add, but a work around is needed for this to work smoothly with the web app. Currently, we estimate that no more than around 10,000 entries can be accommodated without very significantly delaying the queries.
      
%==================================
	\section{Comparing stellar parameters and abundances from multiple literature studies for three metal-poor stars}\label{comparison}
    
		There are a few stars that are regularly used as reference metal-poor stars in abundance studies. In addition, there are many stars that have been analyzed by multiple groups. In total, \textit{JINAbase} contains 437 stars with multiple entries. Each study naturally reports slightly different stellar parameters and chemical abundances, due to the different methods and tools employed. We now assess these systematic differences for the three most well-studied metal-poor stars. HD122563, a cool red giant, has 28 entries in \textit{JINAbase}, HD140283, a subgiant, has 21 entries, and G64$-$12, a main-sequence star, has 16 entries. Furthermore, the evolutionary status of the stars broadly explains the number of reported abundances per study. The red giant star with the intrinsically strongest lines has the largest number of individually reported abundances by all studies. We use this to aid in identifying systematic differences between different studies. On the contrary, for the other two much warmer stars with intrinsically weaker lines, a number of chemical elements are reported only in 1-2 studies.
		
\begin{figure*}[t!]
			\centering
            \epsscale{0.8}
			\plotone{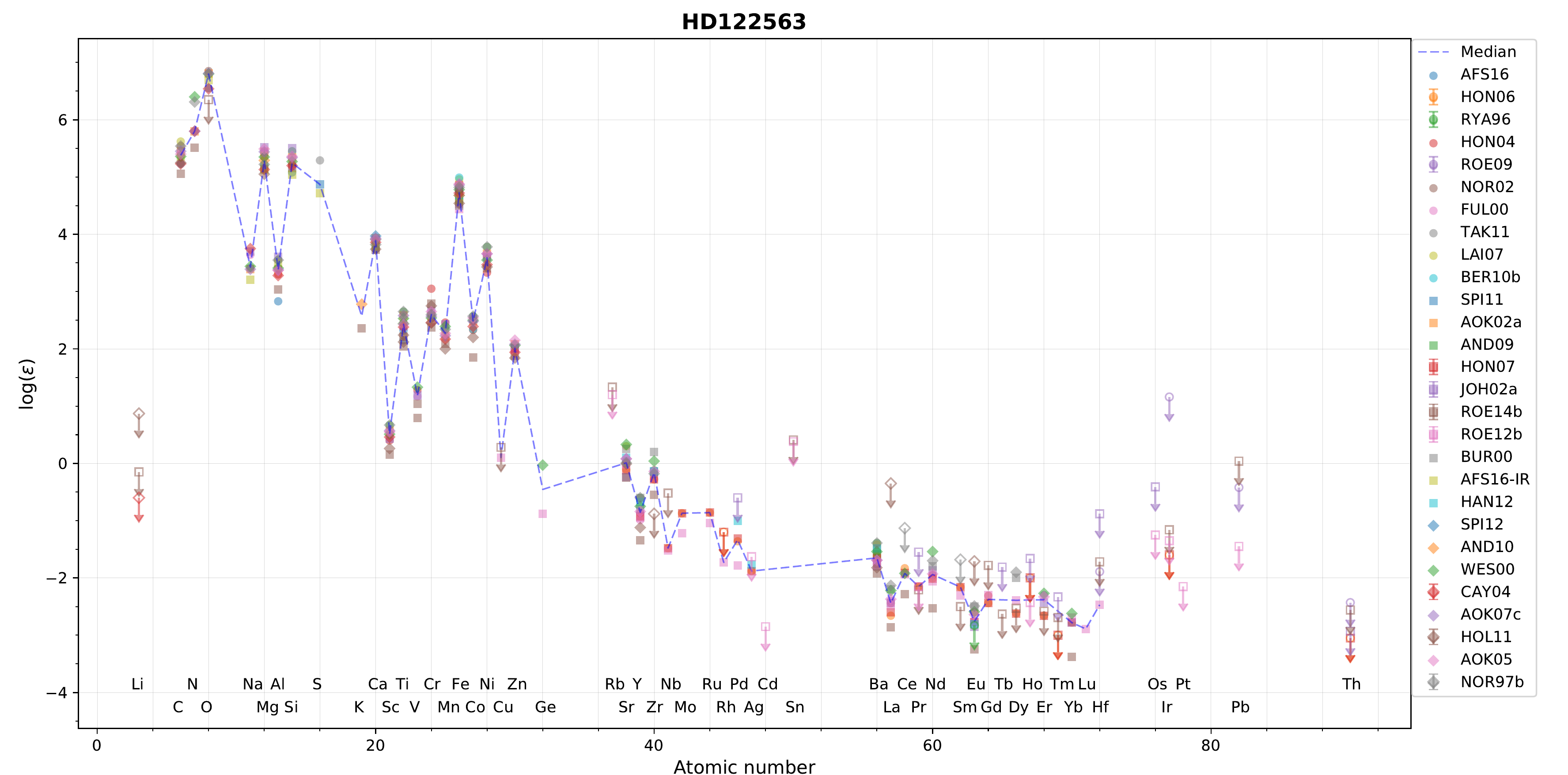}\\
           	\plotone{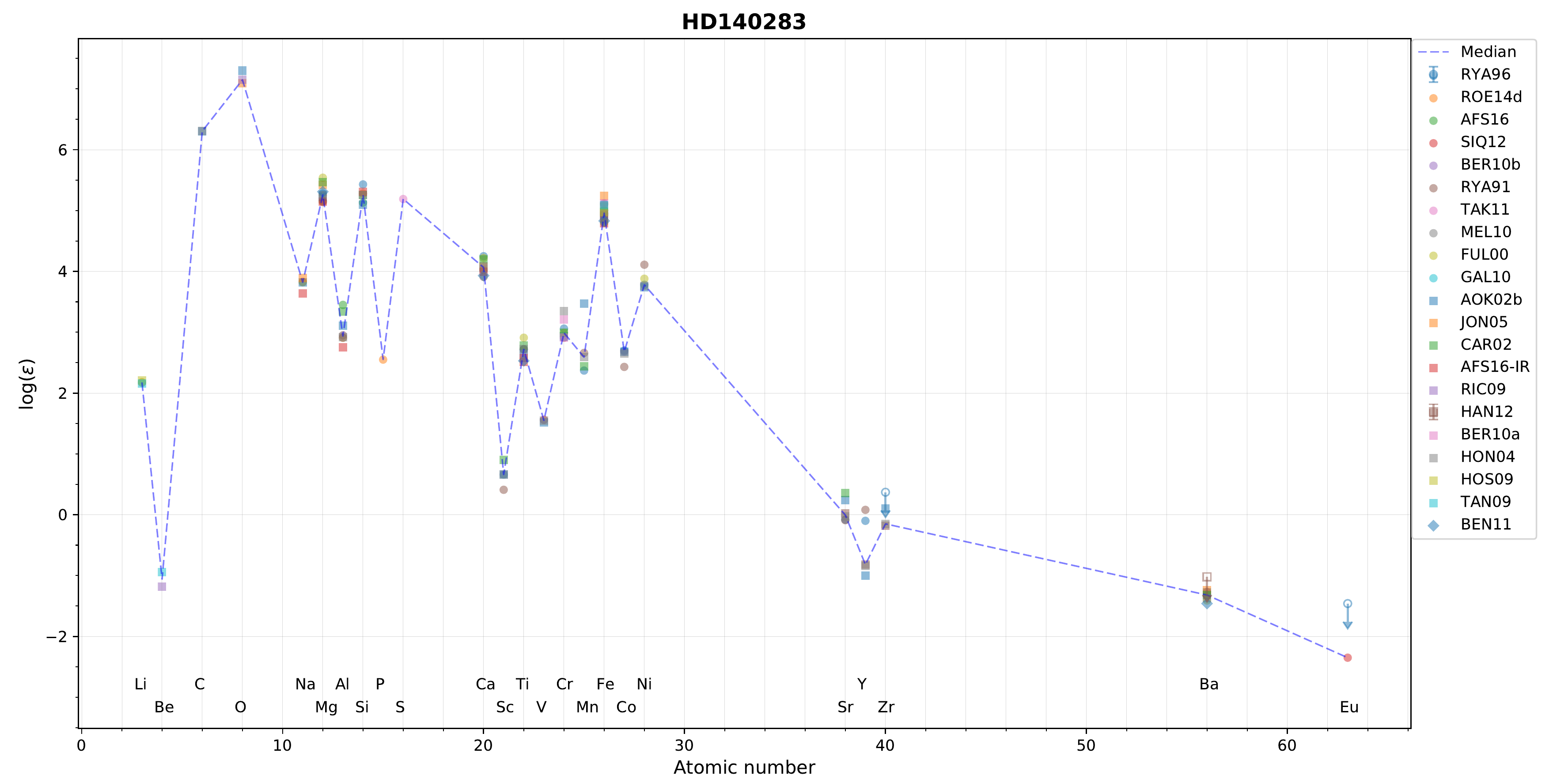}\\
  			\plotone{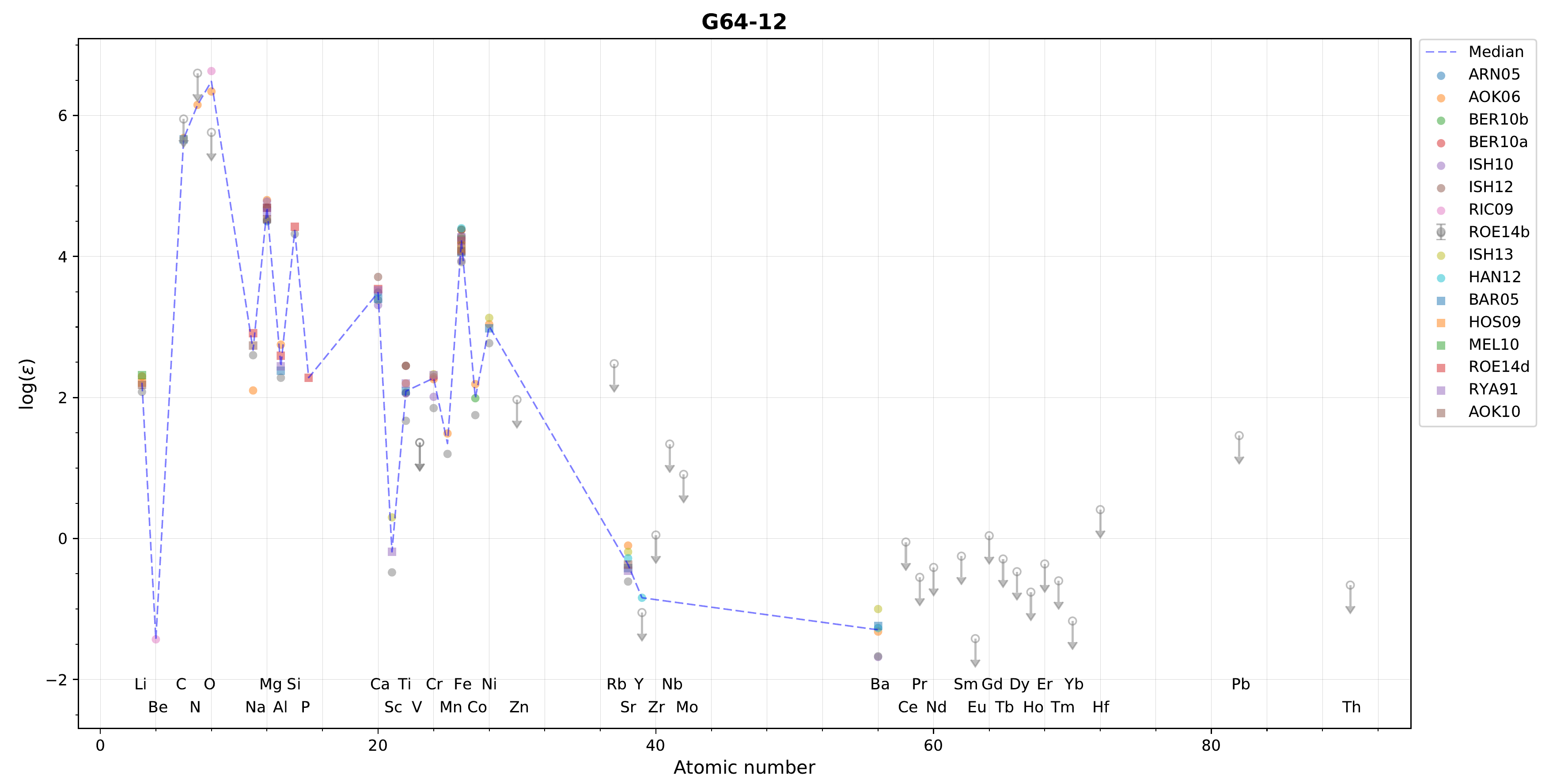}
			\caption{\label{fig:atomicplot} Chemical abundances $\log(\epsilon)$ of available studies in the literature as a function of atomic number for three well-studied metal-poor stars. Results from 28 studies are shown for HD122563 (top), 21 for HD140283 (middle), and 16 for G64$-$12 (bottom). The median for each element is shown as the blue dotted line, upper limits are not included in median calculation. The differences in measurements per chemical elements can be seen from the scatter for each element. References are labeled.}
		\end{figure*}

\begin{figure*}
        	\centering
            \plotone{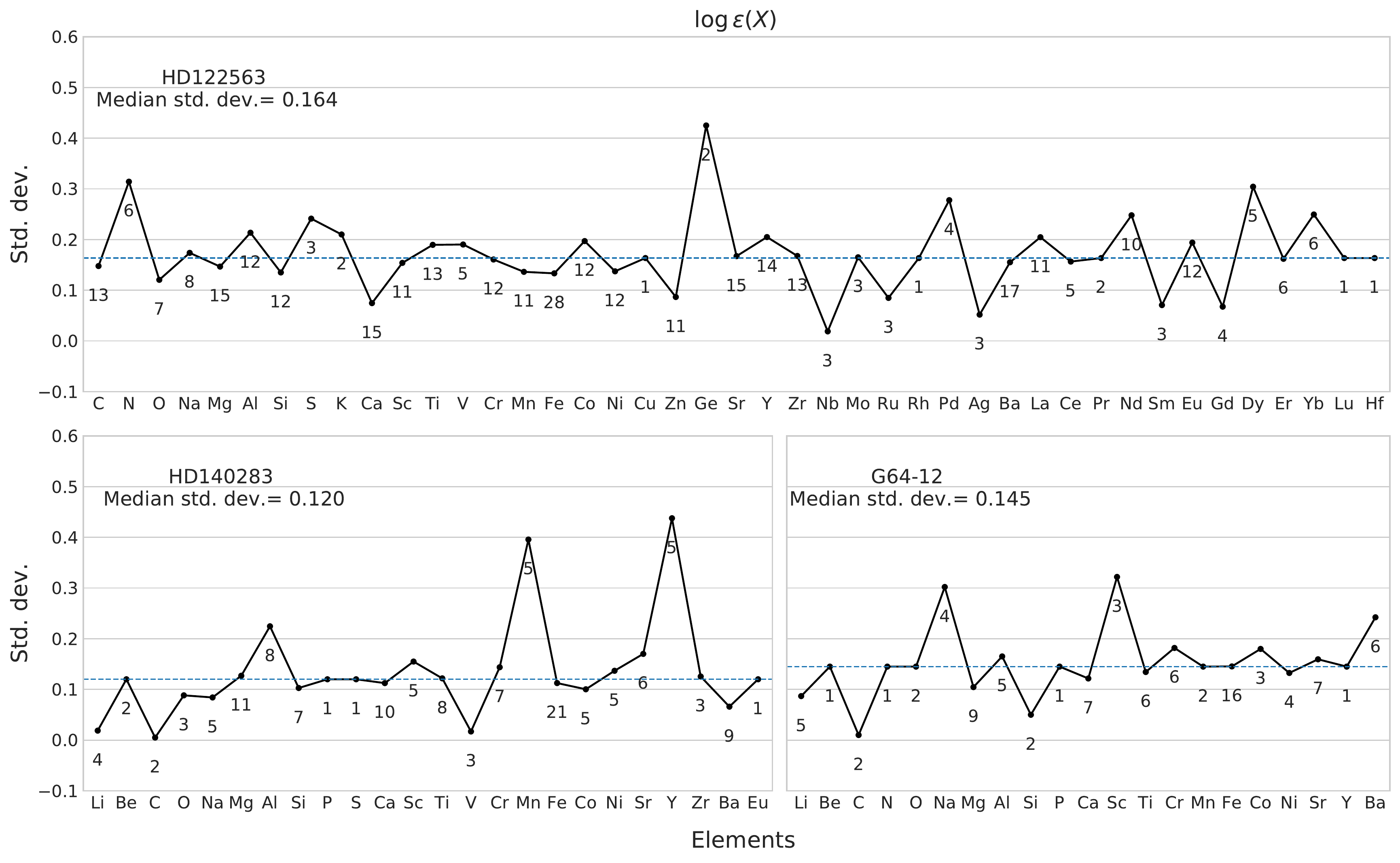}\\
            \plotone{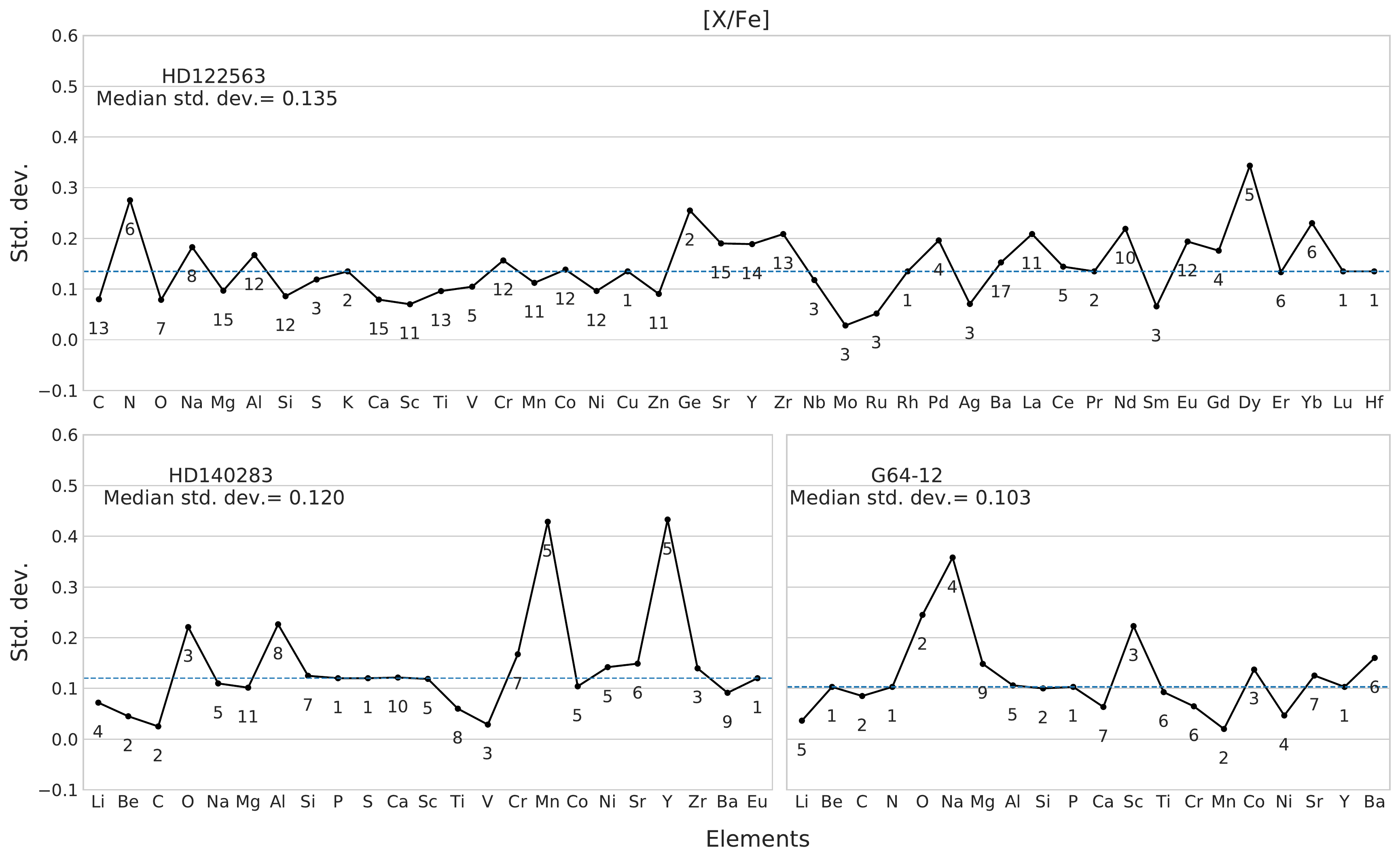}
            \caption{Standard deviation of $\log \epsilon (\rm X)$ abundances (top panel set) and [X/Fe] abundances (bottom panel set) per element from all the different measurements in each HD122563, HD140283 and G64-12. Upper limit measurements are not included. The blue dashed line shows the median standard deviation from all the elements with more than one measurement. For elements with only one measurement, we used the median standard deviation. The number of measurements for each element is shown below the data points.}
			\label{fig:stdplot}
		\end{figure*}

        \smallskip
		\textbf{HD 122563:} In terms of stellar parameters, effective temperature shows a range of 225\,K with a median value of 4600\,K. Surface gravity has a range of 1\,dex with a median of 1.1. Metallicity spans a range of 0.5\,dex with a median of $\mbox{[Fe/H]}=-2.71$. Microturbulence spreads 1.2\,km/s with a median of 2.2\,km/s. 
        Table~\ref{tab:stars} gives more details on the spreads and values in stellar parameters. Figure~\ref{fig:atomicplot} (top panel) shows the various $\log\epsilon(\rm X)$ abundances reported in the literature for HD122563.   
        In Figure~\ref{fig:stdplot} (top row of both panel set), we show the standard deviations of the $\log \epsilon (\rm X)$ and [X/Fe] abundance measurements available for each element. Those vary between 1 and 28 measurements. We also list the number of measurements per element to show the significance of the standard deviation. Upper limits are not included in the standard deviation calculation. The median standard deviation between studies for HD122563 is 0.16\,dex. For elements with only one abundance measurement, we adopt the median standard deviation, for plotting purposes. 
        
        \begin{deluxetable}{l r r r r r}
        \tablecaption{Variations in literature stellar parameters for three well-studied metal-poor stars\label{tab:stars}}
        \tablehead{\colhead{ } & \colhead{Mean} & \colhead{Median} & \colhead{Std} & \colhead{Min} & \colhead{Max}}
        \startdata
          \multicolumn{6}{c}{\textbf{HD122563}}\\
          \hline
          T$_{\rm{eff}}$ & 4568 & 4600 & 65 & 4425 & 4650 \\
          $\log g$ & 1.12 & 1.10 & 0.30 & 0.50 & 1.50 \\
          \mbox{[Fe/H]} & $-$2.77 & $-$2.77 & 0.13 & $-$3.06 & $-$2.51 \\
          v$_{\rm mic}$& 2.20 & 2.20 & 0.31 & 1.70 & 2.90\\
          \hline
          \multicolumn{6}{c}{\textbf{HD140283}}\\
          \hline
          T$_{\rm{eff}}$ & 5737 & 5750 & 50 & 5630 & 5830 \\
          $\log g$ & 3.58 & 3.66 & 0.16 & 3.20 & 3.73 \\
          \mbox{[Fe/H]} & $-$2.52 & $-$2.54 & 0.11 & $-$2.71 & $-$2.26 \\
          v$_{\rm mic}$ & 1.46 & 1.40 & 0.48 & 0.75 & 3.00 \\
          \hline
          \multicolumn{6}{c}{\textbf{G64$-$12}}\\
          \hline
          T$_{\rm{eff}}$ & 6313 & 6333 & 173 & 6030 & 6550 \\
          $\log g$ & 4.17 & 4.20 & 0.35 & 3.58 & 4.68 \\
          \mbox{[Fe/H]} & $-$3.31 & $-$3.27 & 0.14 & $-$3.58 & $-$3.10 \\
          v$_{\rm mic}$ &1.67 & 1.50 & 0.34 & 1.20 & 2.30 \\
        \enddata
        \end{deluxetable}

		\smallskip
		\textbf{HD 140283:} In terms of stellar parameters, effective temperature shows a range of 200\,K with a median value of 5750\,K. Surface gravity has a range of 0.5\,dex with a median of 3.7. Metallicity spans a range of 0.5\,dex with a median of $-2.5$\,dex. Microturbulence spreads a huge 2.25\,km/s with a median of 1.4\,km/s. Table~\ref{tab:stars} gives more details on the spreads and values in stellar parameters. 
        
        Figure~\ref{fig:atomicplot} (middle panel) shows the various $\log\epsilon(\rm X)$ abundances reported in the literature for HD140283. 
        In Figure~\ref{fig:stdplot} (bottom left plots in each panel set), we show the standard deviations of the $\log \epsilon (\rm X)$ and [X/Fe] abundance measurements available for each element. Those vary between 1 and 21 measurements. We also list the number of measurements per element to show the significance of the standard deviation and Upper limits are not included. The median standard deviation for HD140283 is 0.12\,dex. 

\smallskip
		\textbf{G64-12:} In terms of stellar parameters, the effective temperature shows a range of 520\,K with a median value of 6333\,K. Surface gravity has a range of 0.9\,dex with a median of 4.2. Metallicity spans a range of 0.4\,dex with a median of $-3.3$. Table~\ref{tab:stars} gives more details on the spreads and values in the stellar parameters. Microturbulence spreads 1.1\,km/s with a median of 1.5\,km/s. Given the fact that G64$-$12 is extremely metal-poor and fairly warm, only few Fe lines are available. This could explain why the literature values agree to only within $\sim$500\,K if a significant number of these studies used spectroscopic techniques to determine the temperature.
        
        Figure~\ref{fig:atomicplot} (bottom panel) shows the various $\log\epsilon(\rm X)$ abundances reported in the literature for G64$-$12.      
        In Figure~\ref{fig:stdplot} (bottom right plots in each panel set), we show the standard deviations of the $\log \epsilon (\rm X)$ and [X/Fe] abundance measurements (and numbers of measurements) available for each element. Those vary between 1 and 16 measurements. Upper limits are not included. Typical systematic uncertainties between studies for G64$-$12 are $\sim$0.15\,dex based on the median standard deviation.

\smallskip
%conclusion of this section
Despite the many different studies, the effective temperatures are all within $\sim$200\,K for HD122563 and HD140283. Standard deviations are a very reasonable 50-60\,K given that uncertainties of 200\,K are typical across different analysis techniques. For surface gravity, disagreement is large for HD122563 and G64$-$12 on the order of $\sim$1\,dex and moderate for HD140283 (0.5\,dex). Gravity is notoriously difficult to determine (e.g., \citealt{jofre14, heiter15}) which is reflected in these numbers. Metallicity [Fe/H] varies moderately, at the 0.4-0.5\,dex level and likely mainly a result of the uncertainties in the effective temperatures and microturbulence. With differences of more than 1\,km/s, microtubulences wildly disagree between studies which is somewhat concerning.

What is very encouraging, though, is the overall good agreement between all available studies when it comes to the chemical abundance measurements. Average standard deviations between studies for all elements are around 0.15 to 0.20\,dex. These can be regarded as robust general systematic uncertainties on elemental abundances and should be used in various model comparisons. They also agree with typical standard deviations of iron abundances based on line-by-line measurements.

Broadly speaking, typical systematic uncertainties in the analysis of metal-poor stars arising from different analysis methods are thus,
\begin{itemize}
  \item[] $\sigma$\,(T$_{\rm{eff}}$) = 65\,K,\hfill$\sigma$\,($\log g$) = 0.3\,dex,\hfill$\sigma$\,(v$_{\rm{mic}}$) = 0.34\,km/s,\hfill\null
  \item[] $\sigma$\,([Fe/H]) = 0.13\,dex,\hfill$\sigma$\,($\log \epsilon(\rm X)$)=0.15\,dex, and\hfill$\sigma$\,([X/Fe]) = 0.12\,dex.\hfill\null
\end{itemize}

These values are the medians of the standard deviation values; see also Table~\ref{tab:stars} and Fig~\ref{fig:stdplot}.\\

To put these number in perspective, \citep{SMI14} present a statistical analysis of $\sim1300$ FGK-type stars. Their approach is similar to what we tested above. The stars were analyzed by different groups following different methodologies. They then present systematic uncertainties for the stellar parameters and individual chemical abundances for 24 elements. For the effective temperature, systematic differences are 50-100\,K, for $\log g$, they are 0.1-0.25\,dex, and 0.05-0.1\,dex for [Fe/H]. Typical dispersions of the individual chemical abundances range from 0.10- 0.20\,dex. The median uncertainties quoted in \citep{SMI14} are,
\begin{itemize}
  \item[] $\sigma$\,(T$_{\rm{eff}}$) = 55\,K,\hfill$\sigma$\,($\log g$) = 0.13\,dex, and\hfill$\sigma$\,([Fe/H]) = 0.07\,dex\hfill\null%
\end{itemize}

The results found for effective temperature are surprisingly similar. For $\log g$ and [Fe/H], the literature sample has about twice the GESO uncertainties. This can be easily understood given the broad range of analysis, model input choices, and spectra used across all literature studies.

%% 55 K for Teff, 0.13 dex for log g and 0.07 dex for [Fe/H]. Systematic biases are estimated to be between 50-100 K for Teff, 0.10-0.25 dex for log g and 0.05-0.10 dex for [Fe/H]. 
%The typical method-to-method dispersion of the abundances varies between 0.10 and 0.20 dex. 
 
%======================================================
\section{Guide for choosing element abundance samples for model comparisons}\label{samples}

Models of nucleosynthesis and chemical evolution generally rely on comparisons with abundance data from metal-poor stars. With the body of data on these stars being significant by now, it often remains unclear how to choose a suitable sample for model comparisons.

	In the following, we thus comment on various elements, their general availability in abundance studies, usability for model comparisons, caveats about abundance determinations, and uncertainties. This includes the good, the bad, and the ugly from the trenches of the dark art of spectroscopy using mainly optical (and some near-UV and UV) high-resolution spectra. Near infrared studies hold a different set  of secrets. The main purpose is to introduce the interested novice or non-spectroscopists (e.g., students, theorists) to a minimum amount of community wisdom with the aim to assist making useful choices when selecting comparison samples. We critically stress  here that this guide will never be 100\% complete, and it does in no way replace consulting original papers for the many details of each study, and even star.
    
    It should also be noted that many improvements regarding the details of abundance analyses (e.g., NLTE, 3D calculations) have been made in the last few years \citep{gallagher17, nordlander17}. However, these are not yet available for many stars, and whatever has been collected from the literature has likely not (yet) benefited from these advances. We thus focus on common issues that pertain to the majority of elemental abundances measurements in the literature. This includes, unfortunately, that LTE and some NLTE abundances are mixed without any particular labeling.

    Information on isotopic abundances are not discussed further, mainly due to the fact that hardly any are available and usually they have large uncertainties. Elements for which isotopic abundances have been determined in very few stars are Mg, Ba, and Eu. One exception are carbon isotope ratios that are easier to determine and have been measured in many metal-poor stars. However, \textit{JINAbase} has not recorded isotopic abundances.
    
    Finally, uncertainties are further discussed in Section~\ref{comparison} but we note here that bright halo stars for which high quality spectra can be obtained, the uncertainties will be smaller than for dwarf galaxy stars for which barely useful data can just be obtained with the current biggest telescopes.
	
%	As will be highlighted, the source features that abundances are %based on are important and should be considered. [WHAT DOES THIS MEAN?]
    
%	This portion of the paper will for now be handled as a living %document to best serve the JINA community. Please send comments and %request!
	
	\subsection{Lithium}
   
	As stars evolve from the main-sequence to the subgiant branch to the red giant branch, lithium gets increasingly destroyed. The surface abundance thus decreases. With the  convection zone  deepening, lithium from the outermost (observable) layers is transported into hotter, inner regions, where it captures $\alpha$-particles only to then fall apart to beryllium. Hence, lithium can only be detected in warmer stars (up to about the middle of the subgiant branch) that still have thin convection zones.  The corresponding decline in lithium abundance with evolutionary status is well established. 
    
   Near-main-sequence turnoff stars (with no expected surface depletion) show abundances in accordance of what has been termed the ``Spite plateau'', A(Li)$\sim2.3$. (The nomenclature for lithium abundances is different from that of other elements. While other elements are usually given as [X/Fe], lithium is given in the $\log \epsilon$(X) number density format although re-labeled as A(Li), with A standing for ``Abundance''. The reason is that lithium is not produced in fusion like other lighter elements, see more below). At the lowest Fe abundances, there is currently a debate about stars with abundance lower than the Spite Plateau, especially in light of the fact and already the Spite Plateau value is 0.3\,dex lower than what is predicted from standard Big Bang nucleosynthesis. Potential lithium depletion mechanisms in stellar atmospheres are poorly understood, which have lead to this being a decade old problem. 3D and NLTE studies (e.g., \citealt{korn06, lind13}) have provided insight into this problem but could not solve it.
   
    The observed lithium abundances in the most metal-poor stars are thought to reflect the primordial Li abundance. Since then, lithium is only made through cosmic rays and spallation processes and not in stars through the standard fusion processes as all the other lighter elements. Lithium thus does not follow a ``chemical evolution'' like other elements and should not be used for comparisons with e.g., supernova yields or  chemical evolution models.

An interesting alternative use of lithium is when Li-rich giants are found. In a small number of red giants, huge overabundances of lithium are found, in stark contrast to what would be expected from their evolutionary status. Indeed, the reason for the enhancement is that for a short phase up on the giant branch, the Cameron-Fowler beryllium transport mechanism is able to flush Li-rich material to the surface. Then Li-overabundances can be observed. Soon after, convection and dredge up processes will lead to the destruction of the lithium again. It remains unclear if this short lived surface enrichment significantly contributes to the chemical evolution of lithium. It rather provides data to constrain internal stellar evolutionary and mixing processes.

	\subsection{Carbon}
		Carbon is usually determined from molecular features of CH at 4313 and 4323\,{\AA}. In the case of very strong carbon features in the spectrum (due to either a strong carbon enhancement and/or the cool temperature of the star which increases intrinsic line strength), the C$_{2}$ band at $\sim$5130\,{\AA} is also used. It only becomes measurable when the carbon abundance is very high (exact abundances will depend on the temperature of the star, but s-process and i-process stars usually have this occurring). Modeling the line formation of molecules is more complex than that of atomic features. The widespread use of 1D LTE plane-parallel model atmospheres for abundance determination is known to overpredict carbon abundances, especially at low metallicity, compared to 3D hydrodynamically modeled line formation (the temperature structure of the atmosphere is cooler in 3D models which increases line strength and thus reduces the abundance). While corrections, so-called 3D effects, are not yet available for larger samples (each star needs to be modeled individually, e.g., \citealt{gallagher16, gallagher17}), this effect should be kept in mind when comparing with models. Corrections can be as large as 0.7\,dex for extremely metal-poor stars \citep{collet06,nordlander17}. Also, 1D LTE values can be used since relative abundances and abundances spreads can still be assessed. (NLTE studies have not yet been carried out for hydride molecules.)
		
		Carbon abundances are not just affected by our limited capability to model molecular line formation.  A different type of correction that needs to be addressed pertains to the evolutionary status of the star. As the star moves up the red giant branch, carbon gets converted to nitrogen. This results in an in-situ change of the carbon abundance of the observed star. This effect technically renders the star useless for stellar archaeology purposes since its carbon abundance does not reflect the natal cloud anymore, but instead mixing processes in its interior. Fortunately, corrections can be assessed\footnote{They can be obtained here: \\http://www3.nd.edu/$\sim$vplacco/carbon-cor.html} when using as an input the stellar parameters and the 1D LTE carbon abundance. 3D effects are not accounted for in this procedure.
		
		Uncertainties for molecular abundances are usually larger than for other elements, given the spectrum synthesis of an entire band rather than one, clean absorption line. The lack of more than 2-3 bands for any statistics leads to typical uncertainties of about 0.2-0.4\,dex, depending on the data quality and strength of the carbon features.
        %as opposed to 0.1-0.15\,dex for elements with .
	
	\subsection{Nitrogen}
Nitrogen abundances are affected in similar ways as carbon abundances since they are derived from the molecular NH measure at 3360\,{\AA}. Most data do not cover this wavelength regime which explains why only few stars have nitrogen measurements. If the N abundance is not enhanced, chances for detection are decreased as well. Nitrogen is also affected by stellar evolution, it increases as the star moves up the giant branch,
see above discussion on carbon. 3D and NLTE effects equally apply as well, as for carbon. In some cases,  nitrogen is obtained from the CN molecular band at 3880\,{\AA}. In that case, the nitrogen abundance directly depends on the previously determined carbon abundance. This can lead to additional uncertainties and should be kept in mind. Given that only one feature is available in a region where data quality is usually very low, uncertainties are generally large, around 0.3-0.5\,dex.

	\subsection{Oxygen}
		Oxygen is difficult and should only be used with caution. There are three different oxygen indicators, OH molecular lines in the near-UV around 3100\,{\AA}, the forbidden [O] line at 6300\,{\AA} and the O triplet near 7770\,{\AA}. Two of the three indicators suffer modeling issues (3D, NLTE) and the third, largely unaffected, one is a weak feature that is not always measurable. Particularly in metal-poor stars, the forbidden [O] line is generally not detected although it would be       the best  indicator to use. 
        Only OH and O triplet lines are an option but they depend on the O abundance and evolutionary status (i.e., warm stars will likely not show strong features). 
        
        At the lowest iron abundances, only OH lines in the near UV can be measured, if at all. Data quality is usually very low at these wavelengths which prevents detection, although stacking of spectral regions with lines in it can assist with making a combined detection \citep{o_he1327}. OH-based  abundances are affected by 3D effects, in the same ways as carbon as described above. 
		
 The O triplet can still be measured in some extremely metal-poor stars (with $\mbox{[Fe/H]}<-3.0$) but the lines are severely affected by NLTE effects. For O triplet lines, strong NLTE effects that affect each line differently, further complicate their usage \citep{amarsi16b}. Detailed modeling, also in 3D, is required to assess these effects which is still difficult for larger samples.
 
 When putting together a sample of O measurements, original papers should be consulted to construct a sample based on measurements from the same indicator. Even then, corrections should best be taken into account or at least the direction of potential corrections before comparing with model results.
	
	\subsection{Sodium}
		Sodium abundances show a large scatter in halo and dwarf galaxy stars (but a well defined one in globular clusters). In part this might be due to the Na D doublet being easily affected by interstellar Na absorption. In many cases, interstellar absorption is separated from the stellar component by a velocity difference but often no interstellar component is visible. This could be to due no interstellar Na being present or it being aligned with the stellar Na line, thus impacting the stellar Na abundance. Most papers do not report whether or not interstellar absorption was found in the spectrum. This issue may not fully (or at all) explain the Na scatter but it is interesting to keep it in mind that larger Na abundances could be affected this way. An alternative are significant NLTE effects of the order of 0.4\,dex that increase abundances \citep{na_nlte_baum, takeda2003_na_nlte,lind11}. Line-by-line corrections will be needed to assess whether the overall scatter would decrease; in their absence, at least a global correction should be applied before comparing abundances with model results.
	
	\subsection{Magnesium}
		Magnesium is a typical $\alpha$-element. Its abundance is enhanced in halo and dwarf galaxy stars at the $\mbox{[Mg/Fe]}\sim0.4$ level. Nevertheless, a number of halo stars have rather strong Mg (and also Si) enhancements. These stars are likely the result of unusual progenitor supernovae as they significantly deviate from the main trend. 
        
		It can be expected that a variety of oscillator strengths, $\log gf$ values, have been used by the many studies over the last several decades. While the latest measurements by \citep{ALD07} report uncertainties of 9\% in $gf$ values which translates to 0.04\,dex in $\log gf$, previous studies likely used others, some of which were somewhat uncertain, at least for the Mg b lines. This should be kept in mind because it can lead to systematic differences between studies (depending on which values were used) and hence increased scatter. This is in addition to internal scatter between lines, e.g., between the Mg b lines and the blue Mg triplet lines at 3829-3838\,{\AA} which is often larger than the new $\log gf$ uncertainties quoted above.
               
        NLTE effects are similar to those of iron, and yield similar, positive corrections for all lines \citep{mashonkina13, bergemann16}
        %add Ezzeddine et al 2017, in prep later
	
	\subsection{Aluminum and silicon}
		Both elements have only two lines available, one of which (in both cases) is significantly blended with CH and the other one is often too weak to be detected. All four lines are in a lower S/N region (3500-4100\,{\AA}) which often results in large uncertainties. 3D and NLTE have been studied  for Al \citep{nordlander17}. They found corrections to be around 0.4\,dex. For Si, \citet{mashonkina16} and \citep{ezzeddine16a} investigated NLTE effects. We also caution that for Si abundances, there is also a well-established correlation with effective temperature, e.g., \citet{PRE06a}. Ideal samples for comparison with theoretical models might be stars with similar temperatures.

	\subsection{Potassium}
		Potassium is not detected in many metal-poor stars down to $\mbox{[Fe/H]}\sim -4.0$. However, often there appear lines at the K line positions which are atmospheric lines that remain in the spectrum if no telluric line removal has been applied (which usually is not necessary when obtaining optical abundances). If these telluric lines were to be mistaken for K lines then the star becomes potassium enhanced. Original papers should be consulted regarding telluric line removal and other comments. (By the way, potassium-rich stars were once discovered in France after the night assistant at the telescope lit a match in the dome which resulted in stars displaying ``potassium flares'' in their spectra. Makes for a good story rather than good model comparison data, although it can be assumed that this issue is not a problem with modern data, and unrelated to the telluric lines blending with stellar K lines.)
	
	\subsection{Calcium}
		In metal-poor stars, calcium is usually determined from Ca\,I lines. The strongest one is at 4226\,{\AA} which can still be measured in the most metal-poor stars. However, this line is also blended with iron which needs to be taken into account. Moreover, it often gives spurious results compared with other Ca\,I lines in LTE which is unfortunately made worse when applying NLTE corrections. Other, weaker Ca\,I lines are thus principally preferable.
        
        The Ca\,II\,K line is much too strong for abundances measurements except in the few cases of the most iron-poor stars. A few other Ca\,II lines are available around 3700\,{\AA} but are not often used because they are too weak. The Ca\,II triplet lines are another option but their abundances are known to be strongly affected by NLTE \citep{mashonkina_ca,db_SPI12}. As is the case for basically all strong lines, NLTE corrections are large and highly line dependent. This prevents simple adjustment to account for NLTE effects. In a heterogeneous sample, it would thus be best to just use weak Ca\,I abundances.
	
	\subsection{Scandium}
		Scandium is detectable in two ionization stages. In metal-poor stars, only Sc\,II is usually available. Some lines are blended with CH. In the case of carbon-enhanced metal-poor stars, Sc may be unreliable due to severe blending. NLTE corrections for metal-poor stars have not yet been determined for Sc.
	
	\subsection{Titanium}
		Titanium is detectable in two ionization stages. In metal-poor stars, mostly Ti\,II is available. More than a dozen lines are usually available. Regarding NLTE, neutral species are always affected, and ionized species only in a minor way. Hence, Ti\,I is affected by NLTE \citep{bergemann11, sitnova16}
        %xx add ezzeddine et al. 2017 in prep later
        but corrections are similar for all lines and generally well-behaved. Ti\,II is hardly affected and should thus be chosen for abundance comparisons. 
	
	\subsection{Vanadium}
		Vanadium is not (yet) an extensively studied element due to missing good oscillator strength until recently \citep{wood14, lawler14}. Recent works have yielded some abundances for metal-poor stars with $\mbox{[Fe/H]}<-2.5$  but V lines remain undetectable in the most metal-poor stars. All lines are below 4200\,{\AA} which makes detections more difficult.
	
	\subsection{Chromium and manganese}
		Chromium and manganese are usually detectable in two ionization stages in metal-poor stars. The Cr\,I abundances typically are about a few tenths of a dex lower than Cr\,II values. For Mn, Mn\,I lines with higher excitation potential largely agree with the Mn\,II lines when the Mn\,I triplet lines at 4030\,{\AA} are excluded. Original papers need to be consulted to learn about which lines have been used for a given star.
        
       Both Cr\,I and Mn\,I lines are significantly affected by NLTE by several tenths of a dex, and thus produce larger abundances (e.g., \citealt{db_BER10b,BER08}) especially at low metallicity. Given that Cr\,I and Mn\,I lines are more reliably measured than lines of the ionized species, this is a challenging problem. If possible, abundance from lines of ionized species should be used for comparisons with model results. Otherwise NLTE corrected, where available, values based on Cr\,I and Mn\,I can be used, or otherwise uncorrected abundance with appropriate caution.

	\subsection{Iron}
		Iron is the element with the most absorption lines in stellar spectra of metal-poor stars. Its abundance is also used as a proxy for the overall metallicity, and thus often regarded to be the most important element to measure and characterize a star. Unfortunately, obtaining accurate and precise Fe abundances is a complex and challenging undertaking.
        
        Iron is detectable in two ionization stages. In metal-poor stars, Fe\,I is available, Fe\,II often not especially in warmer or more metal-poor stars (but this also depends on the data quality). Statistically speaking, the Fe\,I abundance can be much more accurately determined than any other abundance given that often more than 150  lines are available, as opposed to $\sim$1-30 lines for the other elements. But Fe\,I is affected by NLTE \citep{mashonkina11,bergemann12, sitnova15, ezzeddine16a, amarsi16}, particularly the most iron-poor stars.  Hence, Fe\,II should principally be chosen for abundance comparisons since it is hardly affected by departures from LTE. However, since the ionization balance between Fe\,I and Fe\,II is often used to determine the star's surface gravity (in LTE), the Fe\,II abundance cannot, in practice, offer an advantage after all. This is also the case if any calibration such as that provided in \citep{FRE13} has been applied.

                What to do? Future stellar parameter determinations will take NLTE fully into account \citep{ezzeddine16b}. This will make stars slightly more metal-rich but would solve the problem. The issue is that these calculations are not yet available for almost all stars. In the meantime, we have to use what is available, keeping in mind the limitations. To minimize systematic uncertainties, Fe\,II should be used if determined \textit{independently} from the surface gravity (original papers will need to be consulted). The next best choice are NLTE corrected Fe\,I abundances. If both of these are not available, LTE Fe\,I values need to be used. \citep{ezzeddine16b} has provided a straightforward calibration for adjusting LTE Fe\,I abundances for NLTE effects. Until NLTE values are available for all stars, this might be an acceptable solution.
     
	\subsection{Cobalt and nickel}
		Cobalt and nickel are often detected in stars with $\mbox{[Fe/H]}<-3.0$. With two exceptions, cobalt lines are located below $\sim$3500\,{\AA} where S/N is usually low which increases the abundance uncertainties. Nickel lines are located below $\sim$3600\,{\AA}, with two exceptions, with similar uncertainties. 
        Ni\,I and Ni\,II lines, when both detected, yield consistent abundances in metal-poor giants and main-sequence turn-off stars (e.g., \citealt{db_ROE16b}; \citealt{db_SNE16}), giving confidence to abundances derived from Ni\,I. NLTE for Co has been studied by \citep{db_ZHA09}. However, nothing is published for Ni yet.

	\subsection{Copper}
    	Cu\,I has been measured in relatively few metal-poor stars but down to $\mbox{[Fe/H]}\sim-4$. The two main optical lines used are, among others, at 5105.54\,{\AA} and 5782.13\,{\AA}. Two resonance lines are present in the near-UV spectral range at 3247\,\AA and 3273\,\AA which actually are strong enough for detection in metal-poor stars. There are also several UV lines \citep{db_ROE14e}. However, \citet{bon10} showed that the near-UV lines are affected by 3D corrections under LTE and do not return abundances consistent with those from the optical lines, or with Cu\,II \citep{db_ROE14e}. NLTE calculations (using a new model atom for Cu) by \citet{AND17} show that the abundances determined from the near-UV and UV lines are consistent but agreement with optical lines may only be reached with 3D-NLTE modeling. As such measurements are yet to be computed, original papers need to be consulted to learn which Cu lines were employed and under which assumptions abundance have been determined.
        %Employing the resonance lines, they show that Cu can be detected down to $\mbox{[Fe/H]}=-4.2$ for cooler giants 
        %abdu:none of the entries for that star have cu in jinabase.   

	\subsection{Zinc}
		Zinc lines can be detected down to $\mbox{[Fe/H]}\sim-4.0$ in red giants \citep{db_CAY04,NIS07,db_ROE14b} (and only up to higher metallicities in warmer stars) in the optical wavelength regime. Upper limits in the most metal-poor stars are unfortunately often meaningless because no strong constraints can be derived, especially for the warmer stars. Because of this effect, Zn upper limits are often not reported in papers. This explains the relatively small sample of stars with Zn measurements. A UV Zn line may offer additional Zn measurements in the future for bright stars that can be observed with the Hubble Space Telescope. Zinc lines are affected by NLTE (\citealt{takeda05}, R. Ezzeddine et al. 2018, in prep).

	\subsection{Neutron-capture element examples}
    
    Neutron-capture elements have the vast majority of lines below 4000\,{\AA} where line blending is significant, and data quality is usually low given the reduced stellar flux in combination with decreased quantum efficiency of the CCD detectors. This means, that it is challenging to obtain spectra that are sufficient for detailed studies. Almost all neutron-capture elements present as ionized species which makes them less sensitive to NLTE effects \citep{mashonkina_sr_nlte,mashonkina_ba_nlte}. Exceptions are e.g., Pd, Ag, Cd, Os, Ir, Pt, Au, and Pb which all present as neutral species. 
    
	We now discuss a few representative neutron-capture elements which are most commonly measured. Strontium and barium are detected in \textit{almost all} metal-poor stars and likely present in \textit{all} although possibly in really small amounts \citep{roederer13}. In the few exceptions (often dwarf galaxy stars), upper limits are usually still meaningful because of the intrinsic strengths of the Sr 4077\,{\AA} and Ba 4554\,{\AA} lines. Due to a lack of blue spectra reaching to $\sim4100$\,{\AA}, strontium is measured in comparably few dwarf galaxy stars. But there are several red barium lines which are measurable in red data of dwarf galaxy stars.

		Europium is only detected in neutron-capture-enriched metal-poor stars, such as r-process stars, s-process or i-process stars. The intrinsically weak line at 3819\,{\AA} line is the strongest available  line but located in the blue part of the spectrum and often difficult to detect. The next strongest line at 4129\,{\AA} is thus usually used to identify neutron-capture element enhanced stars. 
        Overall, this means that upper limits on Eu are often rather meaningless because no strong constraints can be derived. Because of this, Eu upper limits are usually not reported. This leads to small samples of stars with actual Eu measurements, especially among dwarf galaxies.

 	\subsection{Lead}
    
Depending on the type of star, lead can be a straightforward measurement or a huge challenge. In s-process metal-poor stars, Pb abundances are high which makes it possible to clearly detect the only Pb line in the optical at 4057\,{\AA}. It is, however, blended with CH, which needs to be taken into account given the fact that s-process stars always have large carbon-enhancements. In r-process stars, the Pb line is hardly detectable and extremely high data quality of $S/N>300$ preferably in an $R>40,000$ spectrum, is needed to attempt it. CH blending remains an issue especially when it is unclear whether Pb is actually detected. Pb appears as neutral species which means NLTE effects are strong. \citet{mashonkina12} calculated corrections of 0.3-0.5\,dex depending on the stellar parameters and Pb abundance. Pb in s-process stars formed through the operation of the s-process at low-metallicity. Pb in r-process stars can provide confirmation on r-process nucleosynthesis calculations because it is the decay products of thorium and uranium.
       
        \subsection{Thorium and uranium}
Thorium has several heavily blended absorption lines of which the 4019\,{\AA} is the best one. It is particularly blended with a $^{13}$CH feature. Most reported Th abundances are based on this line. Uranium is only a tiny blend in the wing of a strong iron line. The detection is very difficult and requires extremely high quality data ($R>40,000$, $S/N \sim300$) to keep observational uncertainties low. NLTE effects are calculated in \citep{mashonkina12}. Th and U measurements are of great interest for carrying out cosmo-chronometry, given their radioactive nature and long half lifes of 4.7 and 14 billion years. Cool giants with the largest overabundances in r-process elements are the most suitable stars to attempt especially a U measurement for age dating purposes. 
Ages very sensitively depend on abundance uncertainties which makes cosmo-chronometry very challenging.

%		\textbf{Retrieve abundances for a group of stars with specific [Ba/Fe] and [Fe/H] values} A small sample of stars show Ba-enhancement, but are not assigned an n-capture key. The upper and lower limits for the x and y axes enables to isolate these stars. The plotted data can be saved as a table formatted text file along with the stellar parameters or other information available in the database. 	

%\section{Examples of JINAbase's use in the literature}

%	This section will be filled in once the JINA community has developed suitable samples. Stay tuned and chime in - send us your projects and we'll help you select the right sample! 

\section{Summary}\label{summary}

We have described a new web application tool, \textit{JINAbase}, to query  literature chemical abundance results of metal-poor stars. Various selection criteria can be used to select very specific samples tailored to a user's need for comparing data with model predictions or to select suitable comparison samples for ongoing abundance studies. Plotting options and options for downloading the associated data are included. Upon registering with the web app, authors can upload their own results which will hopefully contribute to the body of data staying up to date. 
    
\acknowledgments We thank the JINA community for repeatedly voicing their need for a queryable abundance compilation like the one developed here, to maximize options and opportunities for comparing observational data with theoretical model results. We thus thank Benoit Cote, Rana Ezzeddine, Brendan Griffen, Falk Herwig, Alexander Ji, Ian Roederer, Charli Sakari, Hendrik Schatz, and Frank Timmes for helpful discussions and comments on the manuscript. 
%xx DO add others
A.A. acknowledges support from PHY 14-30152; Physics Frontier Center/JINA Center for the Evolution of the Elements (JINA-CEE), awarded by the US National Science Foundation. This research has made use of NASA's Astrophysics Data System. This research has made use of the SIMBAD database, operated at CDS, Strasbourg, France. 

\software{Python \citep{python}, Flask \citep{flask}, Bokeh \citep{bokeh}, SQLAlchemy \citep{sqlalchemy}, Pandas \citep{pandas}, Numpy \citep{numpy}, JINAbase.}

\bibliographystyle{aasjournal}
\bibliography{all_bibtex}

\appendix
\section{Studies included in \textit{JINAbase} (as of July 2016).}
Additional results will be included regularly in the future to keep the \textit{JINAbase} content up to date. Currently, the included studies are:
\citet{db_AFS16,db_ALL12,db_AND09,db_AND10,db_AOK01,db_AOK02a,db_AOK02b,db_AOK02c,db_AOK02d,db_AOK05,db_AOK06,db_AOK07a,db_AOK07b,db_AOK07c,db_AOK08,db_AOK09,db_AOK10,db_AOK12,db_AOK13,db_AOK14,db_ARN05,db_BAB05,db_BAR05,db_BEH10,db_BEN11,db_BER10a,db_BER10b,db_BON09,db_BON12,db_BUR00,db_CAF11a,db_CAF11b,db_CAF13,db_CAL14,db_CAR02,db_CAS15,db_CAY04,db_CHR04,db_COH03,db_COH04,db_COH06,db_COH07,db_COH08,db_COH09,db_COH13,db_COL06,db_COW02,db_CUI13,db_FEL09,db_FRA16, HE1327_Nature, o_he1327, db_FRE07a,db_FRE07b,db_FRE08,db_FRE10a,db_FRE10b,db_FRE14,db_FRE15,db_FRE16,db_FUL00,db_FUL04,db_GAL10,db_GAR09,db_GEI05,db_GIL13, db_HAN11, db_HAN12, db_HAN14, db_HAN15, db_HAY09, db_HIL02, db_HOL11, db_HOL15, db_HON04, db_HON06, db_HON07, db_HON11a, db_HON11b, db_HOS09, db_HOW15, db_HOW16, db_ISH10, db_ISH12, db_ISH13, db_ISH14, db_ITO09, db_ITO13, db_IVA03, db_IVA05, db_IVA06,db_JAB15, db_JAC15, db_JI16a, db_JI16b, db_JOH02a, db_JOH02b, db_JOH04, db_JON05, db_JON06, db_KEL14, db_KEN14,db_KIR15, db_KOC08, db_KOC15, db_LAI07, db_LAI08, db_LAI09,db_LAI11, db_LI13, db_LI15a, db_LI15b, db_LI15c, db_LUC03, db_MAS06, db_MAS10, db_MAS12, db_MAS14, db_MCW95, db_MCW98, db_MEL10, db_MEL16, db_NOR00, db_NOR01, db_NOR02, db_NOR07, db_NOR10, db_NOR12, db_NOR97a, db_NOR97b, db_NOR97c, db_PLA13, db_PLA14a, db_PLA14c, db_PLA15a, db_PLA15b, db_PRE00, db_PRE01, db_PRE06, db_REN12, db_RIC09, db_ROE08, db_ROE09, db_ROE10, db_ROE12a, db_ROE12b, db_ROE14a, db_ROE14b, db_ROE14c, db_ROE14d, db_ROE14e, db_ROE16b, db_ROE16c, db_RUC11, db_RYA91, db_RYA96, db_SAI09, db_SCH07, db_SHE01, db_SHE03,db_SHE13, db_SIM10,db_SIM15, db_SIQ12, db_SIQ14, db_SIV04, db_SIV06,db_STA13, db_SKU15, db_SMI09, db_SNE03, db_SNE16, db_SPI00, db_SPI11, db_SPI12, db_SPI13, db_SPI14, db_SUS16,db_TAF10, db_TAK11, db_TAN09,db_VEN12, db_WES00, db_YON13, db_ZAC98, db_ZHA09}


\end{document}